\documentclass[sigconf]{acmart}

\usepackage{mathtools}
\usepackage{placeins}
\usepackage{latexsym}
\usepackage{tcolorbox}
\usepackage{parcolumns}
\usepackage{booktabs}
\usepackage{tabularx}
\usepackage{ragged2e}
\usepackage{multirow} 
\usepackage{colortbl}
\usepackage{xcolor}
\usepackage{enumitem}
\AtBeginDocument{%
  }

\acmYear{2026}
\setcopyright{cc}
\setcctype{by}
\acmConference[ASSETS '26]{The 28th International ACM SIGACCESS Conference on Computers and Accessibility}{October 25--28, 2026}{Vila Nova de Gaia, Portugal}
\acmBooktitle{The 28th International ACM SIGACCESS Conference on Computers and Accessibility (ASSETS '26), October 25--28, 2026, Vila Nova de Gaia, Portugal}
\acmDOI{10.1145/3797867.3829031}
\acmISBN{979-8-4007-2521-0/2026/10}

\usepackage[english]{babel}
\usepackage{placeins}

\usepackage{amsmath}

\usepackage{graphicx}

\begin{document}

\title[Helping the Helper]{Helping the Helper: LLM-Assisted Problem Articulation for Older Adults Seeking Technology Support}

\author{Hasti Sharifi}
\affiliation{%
  \institution{University of Illinois Chicago}
  \streetaddress{851 S. Morgan}
  \city{Chicago}
  \state{Illinois}
  \country{USA}
  \postcode{60607}
}
\email{hshari3@uic.edu}

\author{Homaira Huda Shomee}
\affiliation{%
  \institution{University of Illinois Chicago}
  \streetaddress{851 S. Morgan}
  \city{Chicago}
  \state{Illinois}
  \country{USA}
  \postcode{60607}
}
\email{hshome2@uic.edu}

\author{Melissa Lamar}
\affiliation{%
  \institution{Rush Alzheimer's Disease Center}
  \streetaddress{1750 W Harrison St}
  \city{Chicago}
  \state{Illinois}
  \country{USA}
  \postcode{60612}
}
\email{melissa_lamar@rush.edu}

\author{Sourav Medya}
\affiliation{%
  \institution{University of Illinois Chicago}
  \streetaddress{851 S. Morgan}
  \city{Chicago}
  \state{Illinois}
  \country{USA}
  \postcode{60607}
}
\email{medya@uic.edu}

\author{Debaleena Chattopadhyay}
\affiliation{%
  \institution{University of Illinois Chicago}
  \streetaddress{851 S. Morgan}
  \city{Chicago}
  \state{Illinois}
  \country{USA}
  \postcode{60607}
}
\email{debchatt@uic.edu}

\renewcommand{\shortauthors}{Sharifi et al.}


\begin{abstract}
Older adults often struggle to articulate technology support needs due to unfamiliar technical terminology and age-related cognitive changes. We explore how large language models (LLMs) can facilitate this problem articulation process. Through a diary study ($n=27$), we identified four communication barriers in older adults' queries: \textit{verbosity}, \textit{incompleteness}, \textit{over-specification}, and \textit{under-specification}. To mitigate these, we developed an LLM pipeline that clarifies context and paraphrases unstructured queries. LLM-rephrased queries significantly improved automated solution accuracy (69\% vs. 35\%). Furthermore, younger adults ($n=48$) acting as technology helpers understood LLM-rephrased queries better (93.7\% vs. 65.8\%) and reported greater ease in providing support. Older adults ($n=34$) also found the resulting solutions highly actionable (94.7\%). Finally, we contribute the first synthetic dataset of older adults' technology assistance queries (STAQ). This work demonstrates how LLMs can improve technology support seeking for older adults by addressing age-related communication barriers.
\end{abstract}



\maketitle

\section{Introduction}

Digital technologies have become essential for accessing information, obtaining services, maintaining social connections, and participating fully in civic and economic life. Consequently, there is a growing need among older adults to engage effectively with these tools \cite{AARPTechTrends2026,Mois_2024_DADL,kim2022exploring}. While a recent survey reports that nearly all U.S. adults aged 50 and older now own at least one device—with smartphone ownership rates comparable to younger adults—device ownership has not inherently translated into broad or sustained use \cite{AARPTechTrends2026,lazar2026interrogating}. To bridge the gap between owning digital devices and using them effectively, older adults often require reliable assistance \cite{sharifi2023senior, pang2021technology, jiang2026help}.

While limited technology uptake can reflect personal preferences \cite{knowles2018wisdom, sakhnini2022review}, many older adults cite the lack of high-quality technology support as a primary barrier to continued use \cite{AARPTechTrends2026, sharifi2023senior}. This need for effective technology support is compounded by several age-related factors. Many older adults encounter contemporary digital systems later in life, without the structured learning or informal exposure that younger generations experience through school, work, and social contexts \cite{elder1994time}. Furthermore, age-related cognitive changes can make it more difficult to learn and retain new information \cite{barnard2013learning,pang2021technology,sharifi2023senior}, frequently leading to a lack of digital confidence \cite{an2022understanding,chu2010outcomes}. These challenges and the fluidity of information to be processed by older adults are exacerbated by the fast-evolving nature of software. As applications frequently update and new interaction paradigms emerge, older adults are forced to acquire entirely new mental models or schema—internal representations of how systems function \cite{gilbert1999age}. Forming these new models requires significant cognitive effort, which may tax an already aging working memory capacity \cite{sweller1988cognitive}, particularly during moments of frustration or urgency (e.g., making repeated unsuccessful attempts \cite{yu2023history} or rushing to cancel a rideshare before a fee is charged \cite{sharifi2023senior}). Consequently, even minor system changes can disrupt an older adult's ability to navigate a familiar tool or explain what went wrong.

When these disruptions occur, older adults typically prefer to seek help from friends, family members, or trusted community resources \cite{an2023weisst,kim2016acceptance,mendel2021exploratory,sharifi2023cross,sharifi2023senior, fang2026guideme}. Alternatively, some, with higher digital skills, attempt to resolve issues independently by turning to search engines or artificial intelligence (AI)-powered chatbots \cite{khurana2024and,sharifi2023senior}. In either scenario, whether interacting with a human helper or an automated system, the first step toward resolving a technology problem is being able to communicate it clearly. However, articulating a technology problem requires more than simply stating that an error occurred; it demands accessing specific vocabulary, applying a mental model of the system, and doing so under cognitive constraints \cite{sharifi2025older}.

Age-related changes in processing speed and working memory capacity can complicate this task \cite{bryan1997verbal,gordon2018older,levitt2006processing,nittrouer2016verbal}. For example, reductions in these cognitive domains that naturally occur with aging may make real-time problem-solving difficult and the articulation of the underlying issue much more challenging \cite{kim2021quantitative}. Furthermore, while older adults often maintain broader general vocabulary knowledge than their younger counterparts \cite{bryan1997verbal}, this advantage rarely extends to technology-related terms that are typically acquired later in life. Moreover, common cognitive conditions such as mild cognitive impairment (MCI) can further impair cognitive abilities beyond expected age-related alterations, making it significantly more difficult to learn a new skill, adapt to system changes, or articulate questions to acquire appropriate solutions \cite{demetriou2017mild,vadikolias2012mild}.


Prior research highlights that older adults frequently experience uncertainty and a lack of confidence when attempting to phrase technology questions effectively \cite{sharifi2023senior, sharifi2023cross, fang2026guideme}. This uncertainty makes technology support problematic on both ends; technology helpers frequently report difficulty understanding older adults’ problem descriptions, particularly during remote or asynchronous communication \cite{sharifi2023cross}. Despite these documented challenges, most AI-powered support systems assume older adults can pose well-formed queries and anchor them to a specific interface or interaction context \cite{gao2024easyask,wang2023enabling,wu2024atlas}. Consequently, there remains a critical gap between how systems expect users to ask for help and how older adults naturally communicate their technical struggles.

In this paper, we examine how English-speaking, community-dwelling older adults communicate technology problems in unstructured contexts, and how large language models (LLMs) can facilitate the problem articulation process by translating natural descriptions into actionable queries. By reducing the cognitive and verbal burden on older adults, we aim to better align technology support systems with real-world help-seeking practices. Specifically, we contribute: (1) a characterization of organically occurring technology support queries from older adults; (2) a few-shot prompt-chaining approach that clarifies, paraphrases, and generates solutions for these queries; (3) an evaluation of this pipeline with both younger and (4) older adults; and (5) the development of STAQ (pronounced ``stack''), a dataset of synthetic technology assistance queries designed to support the training of age-inclusive AI systems. Our findings underscore the necessity of designing technology support tools that reflect the realities of aging, communication, and cognition within sociotechnical ecosystems.

\section{Background}
To situate our work, we review prior literature across three areas: older adults' help-seeking practices, age-related barriers in technology problem articulation, and recent AI applications in technology support.

\subsection{Older Adults and Technology Support}

The definition of an ``older adult'' is not fixed to a specific chronological age \cite{hutchison2010life, vines2015age}. In the literature, definitions vary widely, ranging from individuals aged 50 and above to those 70 and older \cite{wilson2021using, piper2016understanding}. Aging is often experienced relative to specific environments, objects, or rapid technological changes \cite{brandt2010communities}. In this work, we define older adults as individuals aged 60 or over and adopt an anti-deficit model of aging. This perspective avoids treating older users as a ``problematic other'' or viewing aging as inherently disabling \cite{knowles2021harm}. Instead, it recognizes that older adults are active, discriminating users who evaluate technology based on perceived risk, personal values, and cultural expectations \cite{knowles2018wisdom, sharifi2023cross}. Notably, the need for technology support is not universal among this demographic. Many older adults possess high digital proficiency and frequently act as technology helpers for their peers or other family members \cite{geerts2024exploring, sharifi2023cross, sharifi2023senior}.

However, when older adults do struggle with technology use—for instance, when encountering unfamiliar applications or system updates—effective technology support remains critical to prevent abandonment \cite{sharifi2023senior, sharifi2023cross, pang2021technology, fang2026guideme, yu2020maps}. While early technology acceptance models emphasized facilitating conditions like instruction manuals \cite{barnard2013learning, renaud2008predicting, venkatesh2012consumer}, contemporary research indicates a strong preference for interactive and socially situated learning \cite{pang2021technology, sharifi_development_2024}. During initial onboarding, older adults may rely on professional instructors or structured classes \cite{olsson_warm_2018, geerts2024exploring}. However, during continued use, older adults often turn to informal social networks, seeking step-by-step guidance from trusted friends, family members, or peers \cite{sharifi2023senior, fang2026guideme, sharifi2023cross}, or they engage in independent troubleshooting through trial-and-error, online searches, or AI tools \cite{yu2024reducing, gao2024easyask, fang2026guideme}. In either case, the success of the intervention depends heavily on the older adult's ability to clearly communicate the problem to their technology helpers or support tools.

\subsection{Age-Related Barriers in Technology Problem Articulation}

Although older adults’ technology use and help-seeking behaviors have been extensively studied \cite{barnard2013learning,mendel2021exploratory,sharifi2023senior, pang2021technology, tanprasert2024helpcall}, far less attention has been given to \textit{how} they communicate their technology problems. Articulating any complex problem is an inherently demanding cognitive task. It requires working memory to hold the problem's parameters and executive function to retrieve the precise language to describe it \cite{salthouse1991decomposing, salthouse1996processing}. While older adults typically maintain their general vocabulary (crystallized intelligence), they frequently experience age-related declines in fluid cognitive abilities, such as processing speed and working memory \cite{park2002models, salthouse1996processing}. During language production, these fluid changes manifest as real-time word-finding difficulties, tip-of-the-tongue states, and specific declines in action verbal fluency—the ability to rapidly generate verbs \cite{burke1991tip, kim2021quantitative}. Consequently, when explaining an unfamiliar situation, the high cognitive load of navigating the problem space frequently disrupts efficient word retrieval \cite{sweller1988cognitive}. This cognitive bottleneck can force individuals to substitute precise terminology with generalized phrasing, vague referents, or lengthy narratives, creating a fundamental barrier to technology problem communication.

Rather than formulating precise questions, older adults frequently produce ambiguous queries that lack necessary context or rely on vague descriptions of symptoms \cite{fang2026guideme, sharifi2025older}. The cognitive burden of translating a system error into an actionable question can be so high that older adults may struggle silently or abandon tasks altogether, a challenge that has prompted recent research into detecting user frustration implicitly through interaction logs rather than relying on explicit queries \cite{jiang2026help}. Recent reflexive research also suggests that difficulties in problem articulation should not be viewed merely as individual cognitive deficits; they are symptoms of rapidly evolving sociotechnical systems that regularly outpace the mental models of all users, but particularly affect those without continuous technology exposure \cite{lazar2026interrogating}. In structured user studies with well-defined tasks and think-aloud protocols, prior work has found that older adults tend to use specific question words, such as “what,” “why,” and “how” when encountering interaction issues \cite{fan2021older} and ask distinct types of questions during help-seeking from technology helpers \cite{yu2023history}.

A limitation across prior studies examining these communication barriers, is their reliance on specific applications, structured tasks, and controlled settings \cite{gao2024easyask,wang2023enabling,wu2024atlas,yu2023history, fang2026guideme, yu2024reducing}. In organic contexts, problem communication may be significantly more unstructured, lacking in context, and far removed from the precise terminology expected by automated technology support tools or technology helpers.

\subsection{Modern Support Tools using Foundation Models}

The emergence of large language models (LLMs) and vision-language models (VLMs) has shifted technology support from static knowledge bases to dynamic, conversational interventions. Contemporary systems increasingly leverage these foundation models across three primary paradigms: automated customer support \cite{wang2025ecom, su2025llm}, tool learning—where models autonomously execute software tasks \cite{qin2024tool}—and conversational assistants designed to guide software navigation \cite{khurana2024and, khurana2025me}. To ground these conversational assistants in the user's specific context, developers employ different methods for parsing graphical user interfaces (GUIs). Some tools extract the application's underlying view hierarchy and convert it into a structural representation (e.g., HTML) \cite{wang2023enabling}, whereas others rely on VLMs to directly encode the visual and structural elements of the screen \cite{fang2026guideme, wu2024atlas}.

In technology support tools designed specifically for older adults, prior work has leveraged LLMs to mitigate visual and cognitive load, such as assisting users in finding a particular GUI element or remembering a list of task steps. Techniques like dynamically reducing an interface's search space \cite{yu2024reducing} demonstrate how intelligent systems can simplify the navigation of complex applications. Recent systems such as EasyAsk \cite{gao2024easyask} and GuideMe \cite{fang2026guideme} use foundation models to generate contextual, in-app troubleshooting guidance for independent use. EasyAsk employs a multimodal approach, combining voice queries and physical screen touches with an application's underlying view hierarchy to generate interactive tutorials \cite{gao2024easyask}. GuideMe uses VLMs to process raw screenshots alongside natural language queries, generating step-by-step visual instructions overlaid directly on the interface \cite{fang2026guideme}. However, the effectiveness of these assistants depends strictly on the availability of rich contextual inputs—specifically, direct access to the system's screen state or physical interaction with the interface \cite{gao2024easyask, fang2026guideme}.

Despite these advancements, the burden of problem articulation remains a critical bottleneck. Relying on granular screen access \cite{gao2024easyask} or manual screenshots \cite{fang2026guideme} introduces additional effort and privacy concerns for older users. Even when AI generates accurate instructions, non-expert users may struggle to map that text back to the interface \cite{khurana2024and}, yet strongly prefer guided, ``do-it-with-me'' assistance over completely autonomous execution \cite{khurana2025me}. Consequently, while foundation models offer unprecedented troubleshooting capabilities, their practical effectiveness is fundamentally limited by the initial problem articulation phase. If older adults cannot effectively communicate their contextual intent, the resulting guidance—whether AI-generated or human-provided—will inevitably fail to align with their actual needs.

\section{How Older Adults Communicate Technology Problems}

In this section, we examine how English-speaking, community-dwelling older adults describe technology-related problems as they naturally occur in their everyday life. We first detail the methodology of our formative diary study, and then present a qualitative analysis of the unstructured queries collected from our participants.

\newcolumntype{L}{>{\raggedright\arraybackslash}X} 
\newcolumntype{R}{>{\raggedleft\arraybackslash}X}  
\newcolumntype{P}[1]{>{\raggedright\arraybackslash}p{#1}} 
\newcolumntype{K}[1]{>{\raggedleft\arraybackslash}p{#1}}  

\begin{table*}[t]
\centering
\caption{Key Characteristics of Older Adults’ Technology Queries ($N = 57$)}
\label{tab:queries}
\begin{tabularx}{\textwidth}{
    r        
    c
    L   
}
\toprule
\textbf{Characteristic} & \textbf{Count} &\textbf{Example(s)} \\
\midrule
Verbosity & 9 &
``These are the START SCREENS screenshots on [name]’s computer. One for searching, the usual purpose for booting up, and one for Click Bait. My computer combines the two and I do spend some wasted time clicking on the icons. Any way to combine them? Or better, just dispense with the clickbait. And BTW, the favorites are not necessarily the same. She hasn’t complained about hitting Facebook and getting my Facebook account — lately.'' — 80, man

\vspace{0.5em}

``My wife [name] has been advised by trusted source that\ldots Actually, it sounds like something we should both do from time to time\ldots I’m also confused.'' — 70, man
\\
\addlinespace
Over-specification & 12 &
``Argh\ldots I have had this problem for a while and don't even know how or why it happened. I want to watch my wolf utubes with my phone or iPad in the horizontal position. All of a sudden it won't turn when I turn my phone. What happened?!! How do I fix it?!!'' — 66, woman

\vspace{0.5em}

``Need mobile recharge statement for reimbursement but not able to download from app. My son is out of station and he asked for recharge statement for his company for reimbursement.'' — 60, woman
\\

\addlinespace

Under-specification & 13 &
``I'm trying to watch a cooking video on YouTube, but I can't understand what they're saying. Can you help me?'' — 62, woman

\vspace{0.5em}

``I'm really having trouble with my Kindle app. It's been acting weird — the books I read daily aren't syncing between my phone and tablet, and I can't download any new ones. Also, the text is too small, and I have no idea how to fix it.'' — 60, man
\\

\addlinespace

Incompleteness & 24 &
``I come back to phone and it’s blown up so huge I can’t move it. [name] helped me resolve it by just clicking on it. There's gonna be one every day.'' — 61, woman
\\
\bottomrule
\end{tabularx}
\end{table*}





\subsection{Methods}
We conducted an online diary study to examine how older adults communicate technology problems in real-world contexts. Eligible participants were aged 60 or older, used a mobile device (e.g., smartphone or tablet) at least once per week, and were proficient in English. We focused specifically on mobile devices, as they are currently the primary computing platforms used by older adults \cite{AARPTechTrends2026}. Participants were recruited through community flyers, mailing lists, social media platforms, and word-of-mouth. The diary study lasted up to eight weeks, though participants could withdraw at any time. No monetary incentives were provided. After initial contact, a researcher (first author) met with participants to introduce the research study. Participants were asked to message the research team with any mobile device–related technology questions using their preferred communication method (e.g., email, text, or WhatsApp). This approach aligns naturally with asynchronous remote help-seeking, such as via text or email, as well as with independent queries to search engines or AI agents \cite{sharifi2023senior, sharifi_development_2024}. Participants could communicate via text, screenshots, voice messages, or video recordings of their screens. During the first meeting, each participant received both written instructions and video tutorials on how to record audio, screen, and video using their device. They were asked not to share any personal information at any point during the study. To encourage participation, automated reminders were sent every two days, and members of the research team responded with technical assistance as an incentive. All communication, aside from the initial onboarding, occurred asynchronously.

Midway through the study, we observed low engagement, with participants submitting approximately one query per week on average. Follow-up interviews revealed that many older adults typically reached out to family members or trusted technology helpers first and only contacted the research team if the issue remained unresolved. To address this, we expanded recruitment to include adult technology helpers (aged 18 or older) who had previously assisted older adults. These helpers were asked to document asynchronous help interactions by submitting message transcripts, summarizing the issue in their own words, and reporting whether the issue was successfully resolved. As is common in longitudinal research with older adults, not all participants remained engaged for the full eight weeks \cite{abshire2017participant, davies2014improving}.

\subsection{Results}

We collected diary entries from 27 older adults (13 women; \textit{Mdn} age = 66, age range = 60--87). Of these participants, 17 were recruited indirectly through their  technology helpers. Over the study period, the number of queries submitted by each participant ranged from one to ten. All queries were unique; no participant asked about the same issue more than once. Participants reported using a variety of mobile devices, including iPhones, iPads, Android smartphones, and Amazon Fire tablets. 

We collected 57 queries in text format, nine of which included a screenshot. No voice or video recordings were provided. These messages spanned 34 distinct mobile applications. Although all queries addressed different problems, two remained unresolved due to participant attrition. The dataset included both the original query and the full transcript of the asynchronous conversation that followed, either with a technology helper or a researcher. Query length varied substantially, ranging from 5 to 223 words (\textit{Mdn} = 65, \textit{IQR} = 27).

To categorize the \textit{types} of questions, we first applied an existing taxonomy of older adults’ technology-related queries \cite{yu2023history}, which classifies problem articulation into five distinct categories: validation (verifying if an action is correct), directed informational (asking how to execute a specific step), undirected informational (lacking intuition on what UI features to use next), navigational (locating a feature), and conceptual (understanding the system's underlying model). Two researchers independently coded the dataset using a coding reliability thematic approach \cite{braun2021one}, achieving substantial interrater agreement (Cohen’s $\kappa$ = 0.72). Following consensus, the final classification yielded four validation, 14 directed informational, 23 undirected informational, five navigational, and 11 conceptual questions. Although undirected informational queries dominated our corpus, prior research using this taxonomy observed that older adults primarily asked validation questions \cite{yu2023history}. This likely reflects methodological differences. Earlier tasks were synchronous and conducted with a researcher present, likely prompting users to seek reassurance—which was not feasible in our asynchronous, naturalistic setting.


In addition to question type, we conducted a reflexive thematic analysis \cite{braun2021one} to examine how older adults articulated their technology problems. The first author analyzed the queries inductively using open coding, referencing the entire conversation as needed. Then, the codes were discussed and iterated in group data analysis sessions until key themes were identified. To identify the open codes, we grounded our analysis in the aging literature about verbal knowledge, verbal fluency, and age-related changes in fluid and crystallized intelligence. We identified four recurring characteristics in how queries were articulated: excessive detail that obscured the core issue (\textit{verbosity}), inclusion of non-essential specifics (\textit{over-specification}), omission of relevant information (\textit{under-specification}), and a lack of critical context (\textit{incompleteness}).

The technology queries often suffered from verbosity, incorporating superfluous details that delayed a clear understanding of the underlying issue (e.g., \emph{``My wife [name] has been advised by trusted source that...Actually, it sounds like something we should both do from time to time...I'm also confused...''} ---70, man). In contrast, other requests were concise but filled with highly specific situational details, complicating the identification of the root problem (e.g., \emph{``Need mobile recharge statement for reimbursement but not able to download from app. My son is out of station and he asked for recharge statement for his company for reimbursement''} ---60, woman). Queries also frequently lacked granular but critical context---such as whether a user was trying to cancel an ongoing or a scheduled rideshare---or they compounded multiple issues without defining the scope of each sub-problem (e.g., \emph{``I'm really having trouble with my Kindle app. It's been acting weird - the books I read daily aren't syncing between my phone and tablet, and I can't download any new ones. Also, the text is too small, and I have no idea how to fix it.''} ---60, man). In this example, it remains unclear if the text size issue applies to the tablet, the phone, or just the newly downloaded books. Ultimately, the most prevalent characteristic of these queries was the absence of necessary diagnostic information, specifically the device type, operating system, or application in question. Furthermore, participants occasionally used unconventional or metaphorical language that required interpretation—for example, referring to voice commands as `tongue,' describing a locked phone as `power going off,' or calling dark mode `screen going dark.' Table \ref{tab:queries} provides illustrative examples of each characteristic. Note that a query may exhibit multiple characteristics, such as both verbosity and incompleteness.

In summary, our analysis identified several characteristics in older adults’ technology problem articulation that may hinder problem resolution. These patterns, along with occasional use of unconventional vocabulary, point to a need for interventions that support clearer, more effective communication. To address this, we next explore how large language models can be used to reformulate older adults’ queries so that they can be more readily resolved—whether by asking someone, searching online, or interacting with an AI-based support tool.

\section{Reformulating Technology Problems with Large Language Models}

After examining the different characteristics of older adults' technology queries, we set out to reformulate them so that they can be resolved independently, for example, by searching online or asking a chatbot, or communicated effectively to technology helpers. When designing a solution, we chose to address two age-associated characteristics of our dataset: (1) 9/57 entries included a screenshot of the relevant user interface (UI), implying screen-related information is a rarity, and (2) a lack of essential contextual information that the technology helpers had to elicit via follow up questions in order to resolve the problem. These patterns challenge the assumptions of existing question-answering solutions that pertinent UI information is always available, \cite{wang2023enabling, gao2024easyask, wu2024atlas, fang2026guideme}. To address these issues, we next explore how large language models (LLMs) can be used to reformulate older adults’ original queries—making them easier to understand, and ultimately easier to resolve.

\begin{figure*}[ht]
    \centering
    \includegraphics[width=\linewidth]{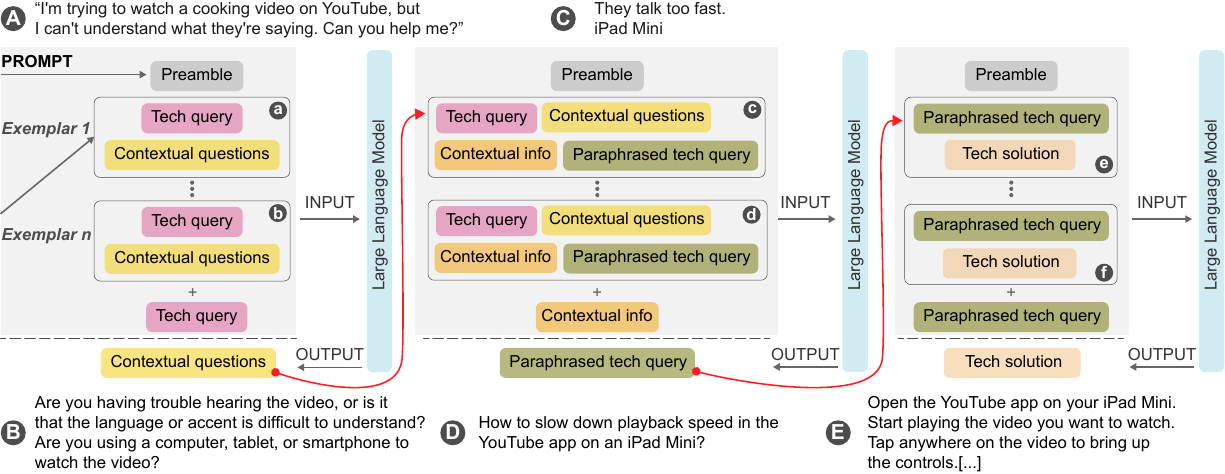}
    \caption{Our \textit{few-shot prompt chaining} approach with large language models operates as follows: First, a query from an older adult is input into the model (A), which is then instructed to generate follow-up contextual questions (B). The model is provided with a few examples (a, b) for guidance. Older adults would then answer these questions to supply contextual information (C). Next, the model takes the original query, contextual questions, and provided contextual information as inputs to generate a paraphrased query (D), with additional examples (c, d) included for reference. Finally, using another set of examples (e, f), the model is asked to generate a solution to the paraphrased query (E).}
\Description[Few-shot prompt chaining for technology support.]{
The figure illustrates a few-shot prompt chaining approach for improving technology support using foundation models. The process begins with an older adult's initial technology-related query (A), which is input into the model. The model then generates contextual follow-up questions (B) with the help of example queries (a, b). The user answers these questions, providing contextual information (C). The model then rephrases the original query using this information (D), incorporating additional reference examples (c, d). Finally, the model generates a technical solution (E) based on the paraphrased query, guided by further examples (e, f).}
    \label{fig:fig1}
\end{figure*}

\subsection{Methods}
We selected two models, GPT-4o \cite{achiam2023gpt} and OS-ATLAS \cite{wu2024atlas}, to develop our approach. GPT-4o is an LLM, and OS-ATLAS is a foundational action model developed for graphical user interface (GUI) grounding and agentic tasks beyond its training distribution. We included OS-ATLAS due to its ability to operate without relying on external APIs and its strength in GUI grounding. GPT-4o was chosen as a general-purpose language model for its strong generalization capabilities and versatility \cite{wu2024atlas}. Compared to other LLMs, such as Gemini, LLaMA, and Claude, GPT-4o demonstrates superior reasoning ability, causal coherence, and overall accuracy \cite{lu2024gpt}. 

For our evaluation pipeline, GPT-4o was specified as OpenAI's Chat Completions API with the model identifier \texttt{gpt-4o}. Because we did not pin a dated snapshot in the initial API configuration, calls used the dynamic production version active during the evaluation window, December 2024--March 2025. OS-ATLAS was specified as \texttt{OS-Copilot/OS-Atlas-Base-7B}, built on the open-source \texttt{Qwen2VLForConditionalGeneration} architecture with a \texttt{Qwen2 -VL-7B} backbone, and evaluated locally using the Hugging Face Transformers library \cite{wolf2019huggingface}.

We used a subset of queries collected in the diary study (described in Section 3) for prompt engineering and model evaluation. Only queries that were understandable without screenshots and had been successfully resolved were included. Two researchers independently reviewed the 51 eligible queries and, when available, their follow-up messages to identify the underlying technology problem. Interrater reliability was high (Cohen’s $\kappa$ = .89). The final dataset consisted of these 51 queries, their associated conversations, and expert annotations of the  problems.

We developed confidence-aware prompts for GPT-4o and OS-ATLAS using an iterative few-shot prompting strategy, with separate iterations tailored to each model’s architecture. Although the final prompts varied slightly, the same three example queries from our dataset were used in the workflow across both models to ensure consistency. Each prompt guided the model through four tasks: (1) analyze the original query to identify missing contextual details; (2) generate up to three follow-up questions to elicit that information; (3) paraphrase the query into one or more concise, complete questions suitable for a Google search; and (4) provide a single, confident solution—defined as at least 90\% confidence for GPT-4o and 80\% for OS-ATLAS (Figure \ref{fig:fig1}).

Throughout the iterations, we tested the prompts using representative examples covering diverse query types, including step-by-step instructions, explanations of technology concepts, and handling of impossible asks. Feedback from these tests was used to refine the prompts, addressing issues such as redundancy in follow-up questions, absence of important details in the paraphrased question, or use of overly technical terminology in the solution. Instructions were added to allow the models to break down complex requests into multiple paraphrased questions. Both models were explicitly directed to prioritize clarity and simplicity in their solutions, avoiding multiple alternatives to ensure usability for older adults. The iterative refinement process ensured that each model's prompt achieved good performance while aligning with the design objectives. To reduce hallucination risk, we implemented confidence thresholds. Although language model “confidence” is a heuristic rather than a statistical measure, these thresholds ensured that solutions were only generated when the model demonstrated sufficient comprehension, thereby improving response reliability. A lower threshold was necessary for OS-ATLAS, because it is a specialized action model rather than a general-purpose LLM; applying the 90\% threshold caused it to over-trigger safety instructions and frequently return ``I don't know.'' The 80\% threshold ultimately provided a better balance between hallucination mitigation and response yield. The final prompts are available in the Appendix \ref{app:a}.


\subsection{Results}
\subsubsection{Model Workflow}
GPT-4o was accessed through the OpenAI API, while OS-ATLAS was run locally. For each of the 48 original queries included in the evaluation (3/51 was used in prompt engineering), we followed a standardized workflow:
\begin{enumerate}
    \item The original query is input into the model, which is then instructed to generate follow-up contextual questions.
    \item Responses to these generated questions, i.e., contextual information, are provided to the model from our corpus collected through the diary study. If no relevant response is available, the model receives “I do not know” as input. Note that our corpus included not only the queries but also the conversations between technology helpers and older adults leading up to problem resolution. Most resolutions required eliciting additional contextual information.
    
    \item The model is subsequently instructed to paraphrase the original query into a concise, complete format suitable for a simple Google search.
    
    \item Finally, the model is asked to provide a single solution to the paraphrased query. Solutions may include step-by-step instructions, conceptual explanations, or a brief rationale for why the query could not be resolved.
\end{enumerate}

\subsubsection{Model Evaluation} 

Model outputs were evaluated for \textit{paraphrasing} and \textit{solution-giving} effectiveness. Paraphrasing effectiveness was assessed based on expert judgment and categorized into three levels: correct, partially correct, and incorrect (see Appendix \ref{app:a} for examples). Partially correct paraphrasing either omitted essential parts of the original query or introduced hallucinated details. Two experts (the first and second authors) independently rated the paraphrasing effectiveness of both models, achieving inter-rater agreement (Cohen’s $\kappa$) of .75 for OS-ATLAS and .85 for GPT-4o. Disagreements were discussed and resolved before final categorization (Table ~\ref{tab:para}). 

Expert evaluation of paraphrasing effectiveness revealed that GPT-4o substantially outperformed OS-ATLAS. GPT-4o produced correct paraphrases for 68.8\% of the queries, compared to only 12.6\% for OS-ATLAS. OS-ATLAS outputs were more frequently rated as partially correct (43.7\%) or incorrect (43.7\%), whereas GPT-4o had far fewer partially correct (25\%) and incorrect (6.2\%) responses. We did not evaluate the paraphrased queries with older adults for two reasons. First, the goal of paraphrasing was to improve the likelihood of obtaining a timely and accurate solution, rather than to assess how older adults perceived the paraphrased queries. Second, the same challenges that led older adults to generate problematic original queries would likely prevent them from accurately judging the effectiveness of the paraphrased queries \cite{wang2019technology, morrison2021older}.

\newcolumntype{L}{>{\raggedright\arraybackslash}X} 
\newcolumntype{R}{>{\raggedleft\arraybackslash}X}  
\newcolumntype{P}[1]{>{\raggedright\arraybackslash}p{#1}} 
\newcolumntype{K}[1]{>{\raggedleft\arraybackslash}p{#1}}  

\begin{table}[h]
\centering
\caption{Expert evaluation of paraphrasing effectiveness (\%)}
\label{tab:para}
\begin{tabularx}{.3\textwidth}{
    r        
    c
    c  
}
\toprule
\textbf{Evaluation} & \textbf{GPT-4o} &\textbf{OS-Atlas} \\
\midrule
Correct & \textbf{68.8} & 12.6 \\ 
    Partially correct & 25 & 43.7  \\
    Incorrect & 6.2 & 43.7 \\

\bottomrule
\end{tabularx}
\end{table}

Solution-giving effectiveness was assessed based on the perceived accuracy of the solution, categorized into three levels: correct, partially correct, and incorrect (see Appendix \ref{app:a}). The ground truth for this evaluation was available because all those technology problems had been resolved during the diary study with older adults. A correct solution for each technology query was created by the first author, who extracted the correct steps from the diary study conversations and structured them following the same instructions given to the models in the prompts. These diary-study resolutions served as a reliable baseline by confirming that each problem was solvable and showing one successful workflow. However, during evaluation, experts assessed whether each model output was functionally accurate, ensuring that technically valid alternative solutions were scored as correct if they resolved the core issue. A comparison of the two models’ effectiveness is presented in Table ~\ref{tab:sol}. In the Google search condition, all correct solutions appeared within the first \textit{five} results, excluding AI-generated content. Only the top 20 Google search results were reviewed to identify a solution. Both models responded with “I do not know” to three queries, one of which was common to both. Additional quantitative evaluation of the model outputs is available in the Appendix \ref{app:a}.

\newcolumntype{L}{>{\raggedright\arraybackslash}X} 
\newcolumntype{R}{>{\raggedleft\arraybackslash}X}  
\newcolumntype{P}[1]{>{\raggedright\arraybackslash}p{#1}} 
\newcolumntype{K}[1]{>{\raggedleft\arraybackslash}p{#1}}  

\begin{table*}[h]
\centering
\caption{Expert evaluation of solution effectiveness (\%)}
\label{tab:sol}
\begin{tabularx}{.55\textwidth}{
    r        
    c
    c  
    ccc
}
\toprule
\textbf{Paraphrasing by} & \textbf{None} &\textbf{GPT-4o} &\textbf{OS-Atlas} &\textbf{OS-Atlas} & \textbf{GPT-4o}\\

\textit{Solution by} & \textit{Google} & \textit{Google} & \textit{Google} & \textit{OS-Atlas} & \textit{GPT-4o} \\ 
    \midrule
    Correct & 35.4 & \textbf{68.8 }& 35.4 & 10.4 & \textbf{68.8} \\
    Partially correct & 2.1 & 22.9 & 10.4 & 10.4 & 14.6 \\
    Incorrect & 62.5 & 8.3 & 54.2 & 79.2 & 16.7 \\
\bottomrule
\end{tabularx}
\end{table*}

Our findings demonstrate that LLMs—particularly GPT-4o—can effectively reformulate older adults’ technology-related queries, even when those queries are vague, imprecise, or lacking critical context. We next examine how LLM-rephrased queries are perceived and interpreted by technology helpers, who play a critical role in resolving older adults’ technology issues.

\section{Evaluation of LLM-Rephrased Queries with Technology Helpers}
\label{sec:4}

Older adults often rely on younger individuals for technology support. However, prior research suggests that younger technology helpers frequently struggle to understand the problem before they can offer effective assistance \cite{gao2024easyask,sharifi2023cross,sharifi2023senior, fang2026guideme}. Ideally, LLM-rephrased queries should help mitigate this communication barrier. To evaluate this possibility, we conducted a controlled experiment with younger adults to assess whether LLM-generated paraphrases of older adults’ technology queries improved perceived understanding, confidence, and ease of interpretation relative to the original queries.

\subsection{Methods}
Adults aged 18 to 50 were recruited to participate in a controlled experiment evaluating LLM-rephrased technology queries. Eligibility criteria included age and demonstration of at least professional working proficiency in English, verified through a brief language screener. No participants from our diary study (Section 3) was part of this evaluation. Participants were recruited via social media platforms and word-of-mouth referrals. No monetary compensation was provided. The experimental materials consisted of query pairs: original queries collected from older adults (as described in Section 3) and their corresponding LLM-reformulated versions generated by GPT-4o using the pipeline outlined in Section 4. To ensure content quality, only query pairs that had received a \textit{correct} paraphrasing rating during expert evaluation were included (69\% of 48 queries). A total of 33 query pairs were included in the study. Each participant rated five of these pairs, which were selected using a Latin square design to ensure balanced and systematic coverage across participants.

Query pairs were presented sequentially—original followed by rephrased—with the order of the pairs randomized for each participant. Participants were not informed which version had been reformulated by an LLM or whether either query was drawn from real users. For each query, participants were first asked whether they understood the query (response options: yes, no, maybe, I don’t know). If they selected any response other than no, they were then asked to rate their confidence in understanding and ease of understanding, both on 5-point Likert scales. 

In addition to these comprehension-related measures, participants completed demographic questions and an adapted version of the Digital Literacy Scale \cite{bayrakci2021digital} (Appendix \ref{app:b}) to assess their own digital proficiency. All procedures were conducted online. Following informed consent and eligibility screening, participants reviewed a randomized set of query pairs. All study materials were administered using the Qualtrics platform.

With query type (original vs. LLM-rephrased) as the independent variable, and comprehension, perceived confidence, and perceived ease of understanding as dependent variables, we tested the following hypotheses:

\textit{H1}: Participants will understand significantly more rephrased queries than original queries.

\textit{H2}: Participants will report significantly greater confidence in understanding rephrased queries than original queries.

\textit{H3}: Participants will find rephrased queries significantly easier to understand than original queries.

\subsection{Results}

\subsubsection{Participants}
Forty-eight participants (\textit{Mdn} age = 28 years, \textit{IQR} = 3.25,8 range = 19--47) completed the study. Each of the 33 unique query pairs was evaluated by between 3 and 6 participants (\textit{Mdn} = 5.0, \textit{IQR} = 1.0). The sample included 25 men, 18 women, 3 participants who identified as genderqueer or gender non-conforming, 1 trans male/trans man, and 1 participant who preferred not to disclose their gender. Participants represented a diverse range of ethnic backgrounds: 19 identified as White, 11 as Asian, 9 as Middle Eastern or North African, 4 as Hispanic/Latino, 2 as Black or African American, 2 as Other, and 1 declined to report their ethnicity. Participants reported high digital literacy (\textit{Mdn} = 4.88, \textit{IQR} = 0.31, on a 5-point scale). Most participants (\textit{n} = 44) indicated either native or full professional proficiency in English. Additionally, a majority (\textit{n} = 31) had experience helping older adults with technology, primarily in a personal capacity; among these, 23 participants (74\%) had been providing technology support for more than five years.

\subsubsection{Hypotheses Testing}
Participants reported understanding the rephrased queries significantly more often (\textit{n} = 149, 93.7\%) than the original queries (\textit{n} = 106, 65.8\%). A McNemar’s test confirmed that this difference was statistically significant, $\chi^2$(1, \textit{N} = 55) = 35.20, \textit{p} $< $.001, with a large effect size ($OR$ = 10.0). Thus, H1 was supported.

\begin{figure*}[t]
    \centering
    \includegraphics[width=\linewidth]{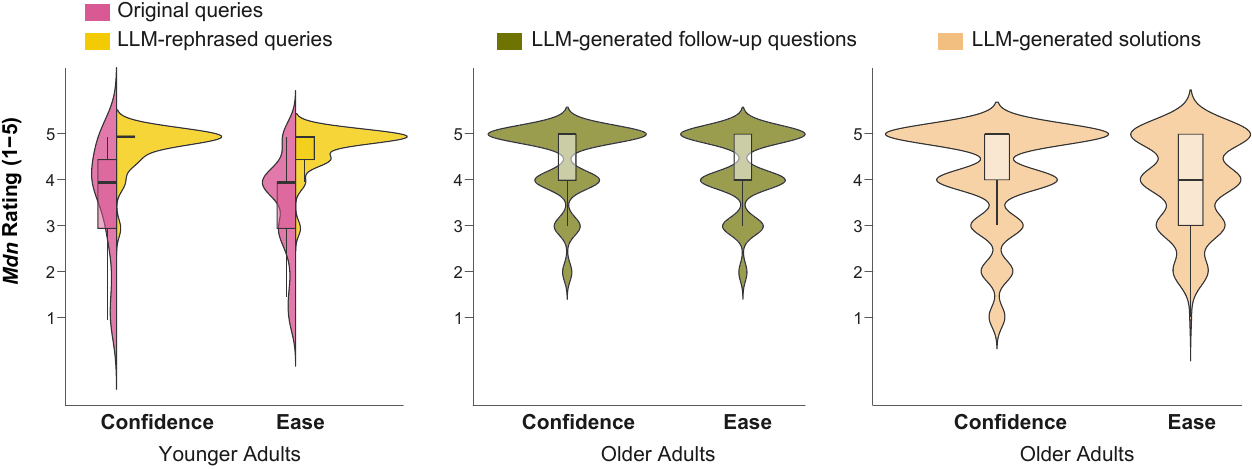}
    \caption{Younger adults reported significantly greater confidence and ease in understanding the LLM-rephrased queries (left). Older adults rated their confidence in responding to LLM-generated follow-up questions as high, but their ease of answering as moderate (center). Similarly, older adults reported high confidence in their ability to follow the solutions, but rated the ease of doing so as moderate (right).}
\Description[Experimental Results]{Three panels of violin plots with overlaid box plots. In all panels the vertical axis is median rating from 1 to 5, and the horizontal axis has two groups: Confidence and Ease.
Left panel is about younger adults. Each violin is split: the left half (pink) shows original queries, the right half (yellow) shows LLM-reshaped queries. For original queries, ratings spread across the full 1–5 scale, with a median of 4 and an interquartile range of about 3 to 4.5 for both confidence and ease. For LLM-reshaped queries, ratings concentrate tightly at the top of the scale: the median is 5 for confidence and close to 5 for ease, with the interquartile range compressed between roughly 4.5 and 5. Almost no mass falls below 3.
Center panel is about older adults, LLM-generated follow-up questions (olive). Both distributions are bimodal, with peaks at 5 and at 4 and a smaller tail toward 2. Confidence has a median of 5 with an interquartile range of 4 to 5. Ease has a median of 4 with the same interquartile range of 4 to 5.
Right panel is about older adults, LLM-generated solutions (peach). Confidence has a median of 5 and an interquartile range of 4 to 5, with the distribution peaking sharply at 5. Ease has a median of 4 and a wider interquartile range of 3 to 5, with visible mass at 3 and a tail extending toward 1.
Across all three panels, the confidence median is at or above the ease median, and the ease distributions are consistently wider.
}
    \label{fig:exp}
\end{figure*}

A Wilcoxon signed-rank test revealed that participants expressed significantly greater confidence in understanding the rephrased queries (\textit{Mdn} = 5.00, \textit{IQR} = 0.00) compared to the original queries (\textit{Mdn} = 4.00, \textit{IQR} = 1.50), \textit{V} = 287.00, \textit{p} $< $ .001, with a medium effect size (\textit{r} = .50), supporting H2 (Figure ~\ref{fig:exp}).

Participants also rated the rephrased queries as significantly easier to understand (\textit{Mdn} = 5.00, \textit{IQR} = 0.50) than the original queries (\textit{Mdn} = 4.00, \textit{IQR} = 1.00), \textit{V} = 395.00, \textit{p} $< $ .001, with a medium effect size (\textit{r} = .56), supporting H3 (Figure ~\ref{fig:exp}).

\subsubsection{Additional Analysis}


Spearman’s rank-order correlations were conducted to examine the relationship between participants’ self-reported digital literacy and understanding, confidence, and ease when responding to the original and rephrased queries. For original queries, digital literacy was weakly positively correlated with understanding, $\rho$ = .12, \textit{p} = .041, and weakly negatively correlated with confidence, $\rho$  = –.12, \textit{p} = .038. No significant correlations were observed for rephrased queries: digital literacy was not significantly associated with understanding, confidence, or ease. These results suggest that participants with higher digital literacy were slightly more likely to understand the original queries but paradoxically felt less confident in their responses.

To further explore the impact of rephrasing, an improvement score was calculated as the difference in understanding between rephrased and original queries. A Spearman correlation revealed a significant negative association between digital literacy and improvement in understanding, $\rho = -.33$, $p = .022$. This indicates that participants with lower digital literacy benefited more from rephrased queries, reporting greater gains in understanding than their more digitally literate counterparts.

We also tested whether prior experience as a technology helper to older adults influenced comprehension. A Mann–Whitney U test indicated no significant difference in comprehension of original queries between technology helpers and non-helpers, \textit{U} = 3243.00, \textit{p} = .193. To determine whether technology helpers and non-helpers reported equivalent understanding of rephrased queries, a Two One-Sided Tests (TOST) procedure was conducted \cite{lakens2017equivalence}. Equivalence bounds were set at ±0.50 standardized units, corresponding to a moderate effect size (Cohen’s \textit{d}). The observed mean difference in understanding was small (\textit{M} difference = 0.08), and both one-sided tests were statistically significant, \textit{p}s = .030 and $<$ .001. These results support the conclusion that technology helpers and non-helpers reported statistically equivalent levels of understanding for rephrased queries, within a moderate equivalence margin.

In this section, we showed that LLM-rephrased queries improved comprehension, confidence, and clarity among younger adults, including those with prior experience supporting older adults with technology. However, communicating the problem is only one part of the support process. In the next section, we assess whether the LLM-generated contextual questions and solutions are understandable and useful to older adults themselves.

\section{Evaluation of LLM-Generated Solutions with Older Adults}

In Section 4, we introduced our few-shot prompt chaining approach (Figure \ref{fig:fig1}), which rephrases the original technology query and generates a corresponding solution. However, the success of this pipeline hinges on two critical factors: first, whether older adults feel comfortable and capable of answering the contextual follow-up questions generated by the model; and second, whether the solutions themselves are understandable, actionable, and elicit confidence. Although expert raters found the GPT-4o-generated solutions to be accurate 69\% of the time, their practical value depends on how well older adults can understand and follow them. In this section, we assess those aspects with older adults.

\subsection{Methods}
We conducted a controlled experiment to evaluate older adults’ perceptions of LLM-generated follow-up questions and solutions. Participants were eligible if they were aged 60 years or older and demonstrated at least limited working proficiency in English, verified through a self-assessment screener. No participants from our diary study (Section 3) was part of this evaluation. Recruitment was conducted via social media platforms and word-of-mouth referrals. No monetary compensation was offered.

Participants evaluated LLM-generated follow-up questions based on three criteria: (1) whether they felt able to answer the question (response options: yes, no, maybe), (2) their confidence in their answer, and (3) the perceived ease of answering it. They also evaluated LLM-generated solutions using a parallel set of measures: (1) whether they felt able to follow the solution (response options: yes, no, maybe), (2) their confidence in following it, and (3) the perceived ease of doing so. If participants selected any response other than no, they were asked to rate their confidence and ease using 5-point Likert scales.

Eighty-five unique follow-up questions and 36 unique solutions generated by GPT-4o were included for evaluation. Only solutions that had received a correct rating from expert evaluators in Section 3 were included to ensure content quality. To personalize the study experience, participants were first asked which mobile devices they currently use (e.g., smartphone, tablet, Kindle). Survey content was then dynamically filtered so that participants only reviewed content relevant to the devices they reported using. Three follow-up and two solutions were not rated because none of our participants reported using the related mobile devices. 

After completing eligibility screening and providing informed consent, participants were shown a randomized subset of follow-up questions and solutions. They then rated each item using the measures described above. All materials were administered online using the Qualtrics platform. At the end of the study, participants completed a brief demographic questionnaire and the 14-item Mobile Device Proficiency Questionnaire (MDPQ-14; range: 14 to 70 \cite{roque2018new}).

\subsection{Results}

\subsubsection{Participants}
Thirty-four older adults (\textit{Mdn} age = 72 years, \textit{IQR} = 10.75, range = 63--91) participated in the study. Of those who provided demographic information, 19 identified as women and 7 as men. The majority of participants identified as White (\textit{n} = 25). Educational attainment varied: seven participants reported holding a high school diploma, three had a 2-year degree, seven held a 3- or 4-year degree, seven had a master’s degree, and three held a doctoral degree. In terms of living arrangements, 18 participants lived independently, four lived with family, and four resided in a senior living community. English proficiency was generally high, with 17 participants reporting full professional proficiency, eight reporting native or bilingual proficiency, five reporting professional working proficiency, and four reporting limited working proficiency. Self-reported mobile device proficiency was moderate to high (\textit{Mdn} = 62.0, \textit{IQR} = 2.15).

Participants rated 82 unique LLM-generated follow-up questions and 34 unique LLM-generated solutions. The number of follow-up questions rated per participant ranged from 1 to 20 (\textit{Mdn} = 16, \textit{IQR} = 3.75), with most participants rating between 14 and 18 items. For the solution items, participants rated between 1 and 12 solutions (\textit{Mdn} = 6, \textit{IQR} = 5.0), with most providing responses for 6 to 10 solutions.

\subsubsection{LLM-Generated Follow-up Questions}
Participants reported high levels of perceived ability to respond to the LLM-generated follow-up questions. Across all ratings, 79.8\% of responses indicated `Yes' and 10.0\% indicated `Maybe' when asked if they could answer the question, suggesting strong engagement with some uncertainty. Confidence in their ability to respond was also high (\textit{Mdn} = 5.0, \textit{IQR} = 1.0), while ease of answering was rated as moderate (\textit{Mdn} = 4.0, \textit{IQR} = 1.0; Figure ~\ref{fig:exp}).

Spearman’s rank-order correlation revealed that mobile device proficiency was positively associated with both confidence ($\rho$ = .40, \textit{p} $<$ .001) and ease of answering ($\rho$ = .56, \textit{p} $<$ .001). There was also a small but statistically significant correlation between digital proficiency and likelihood of responding `Yes' to whether they could answer the question ($\rho$ = .18, \textit{p} = .003), indicating that participants with higher technology proficiency felt more capable of responding the contextual questions.

\subsubsection{LLM-Generated Technology Solutions}
Regarding the LLM-generated solutions, 81.6\% of responses indicated participants could follow the solution, and an additional 13.1\% responded `Maybe.' These results suggest that most participants found the solutions actionable, with a subset expressing partial confidence or uncertainty. Participants reported high confidence in their ability to follow the solutions (\textit{Mdn} = 5.0, \textit{IQR} = 1.0) and generally rated the ease of following the instructions as moderate (\textit{Mdn} = 4.0, \textit{IQR} = 2.0; Figure ~\ref{fig:exp}).

Spearman rank-order correlations were also conducted to examine whether digital proficiency and education level were associated with participants’ confidence, ease, and perceived ability to follow proposed technology solutions. Digital proficiency showed strong, statistically significant associations with all three outcomes: confidence ($\rho$ = .53, \textit{p} $<$ .001), ease ($\rho$= .47, \textit{p} $<$ .001), and ability to follow the solution ($\rho$ = .44, \textit{p} $<$ .001). In contrast, education level was moderately associated with confidence ($\rho$ = .34, \textit{p} $<$ .001) and perceived ability to follow the solution ($\rho$ = .31, \textit{p} $<$ .001), but not significantly related to ease. 

In sum, our results show the promise of LLM-generated query reformulation and technology support in improving older adults’ technology experiences. However, scalable development remains challenging. Most models are trained on data from younger, digitally fluent users, often overlooking the communication styles and cognitive needs of aging populations \cite{wu2024atlas, wang2023enabling}. To address this, next we created a synthetic dataset that reflects these overlooked characteristics.

\section{STAQ: A Synthetic Technology Assistance Queries Dataset for Training Age-Inclusive AI Systems}

Developing equitable AI systems that support older adults requires training data that reflect the realities of aging. Within HCI, there is a rightful consensus that empirical, human-centered data collection cannot be replaced. However, requiring older adults to participate in the exhaustive, high-volume data generation necessary to pre-train or fine-tune modern foundation models is both practically prohibitive and ethically burdensome. Consequently, most large-scale training datasets default to younger demographics, leaving older adults structurally underrepresented. We acknowledge that synthetic data should never replace participatory design or human evaluation; rather, it can serve as a necessary, scalable bridge for model training. By using large language models to systematically amplify empirically observed behaviors—rather than relying on stereotypical assumptions—researchers can simulate underrepresented communication barriers without extracting endless labor from vulnerable populations. To this end, this section introduces STAQ (Synthetic Technology Assistance Queries), pronounced ``stack'': a synthetic training dataset grounded in the real-world, unstructured queries collected during our formative diary study (Section 3). Our dataset is publicly available in the Zenodo repository\footnote{https://doi.org/10.5281/zenodo.21323107}.

\subsection{Dataset Development}
We generated synthetic data using few-shot prompting with GPT-4o, guided by the communication characteristics of older adults’ technology queries identified in Section 3: verbosity, over-specification, under-specification, and incompleteness. Each prompt included three example pairs—an original query exhibiting one of these characteristics and a corresponding expert-generated paraphrase (see Appendix \ref{app:c}). To ensure diversity while maintaining realism, the model was instructed to vary the types of technical issues. To prevent the model from generating ageist stereotypes, we did not instruct the LLM to emulate a generic ``older adult.'' Instead, the generation was strictly anchored in the empirical data, requiring the model to mirror the specific conversational phrasing, expressions of uncertainty, and informal tone observed in our real-world dataset.

To examine linguistic characteristics and validate the realism of the generated text, we analyzed token length and lexical diversity. On average, the synthetic original queries contained $M = 58.89$ tokens, whereas the paraphrases averaged $M = 23.75$ tokens (range: 13 to 174 tokens). We also calculated the type-token ratio (TTR) to assess lexical repetition. In the synthetic dataset, queries written in the older adult style yielded a TTR of 0.858, compared to 0.937 for expert paraphrases. A similar pattern was observed in the real-world dataset (TTR = 0.842 for older adults vs. 0.890 for experts), demonstrating that the synthetic queries successfully mirrored authentic communication patterns.

\subsection{Dataset Composition}

The STAQ dataset comprises 514 instances, distributed across four categories: verbosity ($n = 120$), over-specification ($n = 120$), under-specification ($n = 154$), and incompleteness ($n = 120$). Each instance includes a technology-related query written in the communication style of older adults, paired with an LLM-generated paraphrase that clarifies the intended meaning. Illustrative examples are provided in Table ~\ref{tab:syn}.

\newcolumntype{L}{>{\raggedright\arraybackslash}X} 
\newcolumntype{R}{>{\raggedleft\arraybackslash}X}  
\newcolumntype{P}[1]{>{\raggedright\arraybackslash}p{#1}} 
\newcolumntype{K}[1]{>{\raggedleft\arraybackslash}p{#1}}  

\begin{table*}[t]
\centering
\caption{Example queries from the STAQ dataset and their characteristics.}
\label{tab:syn}
\begin{tabularx}{\textwidth}{
    r        
    P{7cm}
    P{4.8cm} 
}
\toprule
\textbf{Characteristic} & \textbf{Synthetic Query} & \textbf{AI-Rephrased Query} \\
\midrule
Verbosity &
You know, ever since my granddaughter gave me this tablet thingie, I've been having a hard time trying to post photos on my Facebook. Every time I try to post a photo, it just gets stuck or something and then the tablet just freezes up. I can't even turn the darn thing off then. I try pressing the button thing to turn it off, but nothing happens. It used to be easier with my old computer. I've tried waiting it out and it still doesn't work right. Do you think you could help with that? &
Why does my tablet freeze and become unresponsive when I try to post photos on Facebook, and how can I fix it? \\

\addlinespace

Over-specification &
I've been trying to connect to that zoom thing, you know, for the church meeting. But it's been quite a headache. I finally got the app on my computer, but now it's asking for a password or something. I just don't quite understand why or how to get that. Can you help explain? &
How do I join a Zoom meeting and where do I find the password for the meeting? \\

\addlinespace

Under-specification &
I want to send pictures to my grandson but my Gmail is just not doing it. &
How can I successfully send images through email using Gmail? \\

\addlinespace

Incompleteness &
Hey, my Facebook is not working like before. How do I make it normal? &
How can I revert to the previous version or settings of my Facebook application as it is not behaving as expected currently? \\

\bottomrule
\end{tabularx}
\end{table*}

\subsection{Dataset Evaluation}
\begin{figure}[h]
    \centering
    \includegraphics[width=\linewidth]{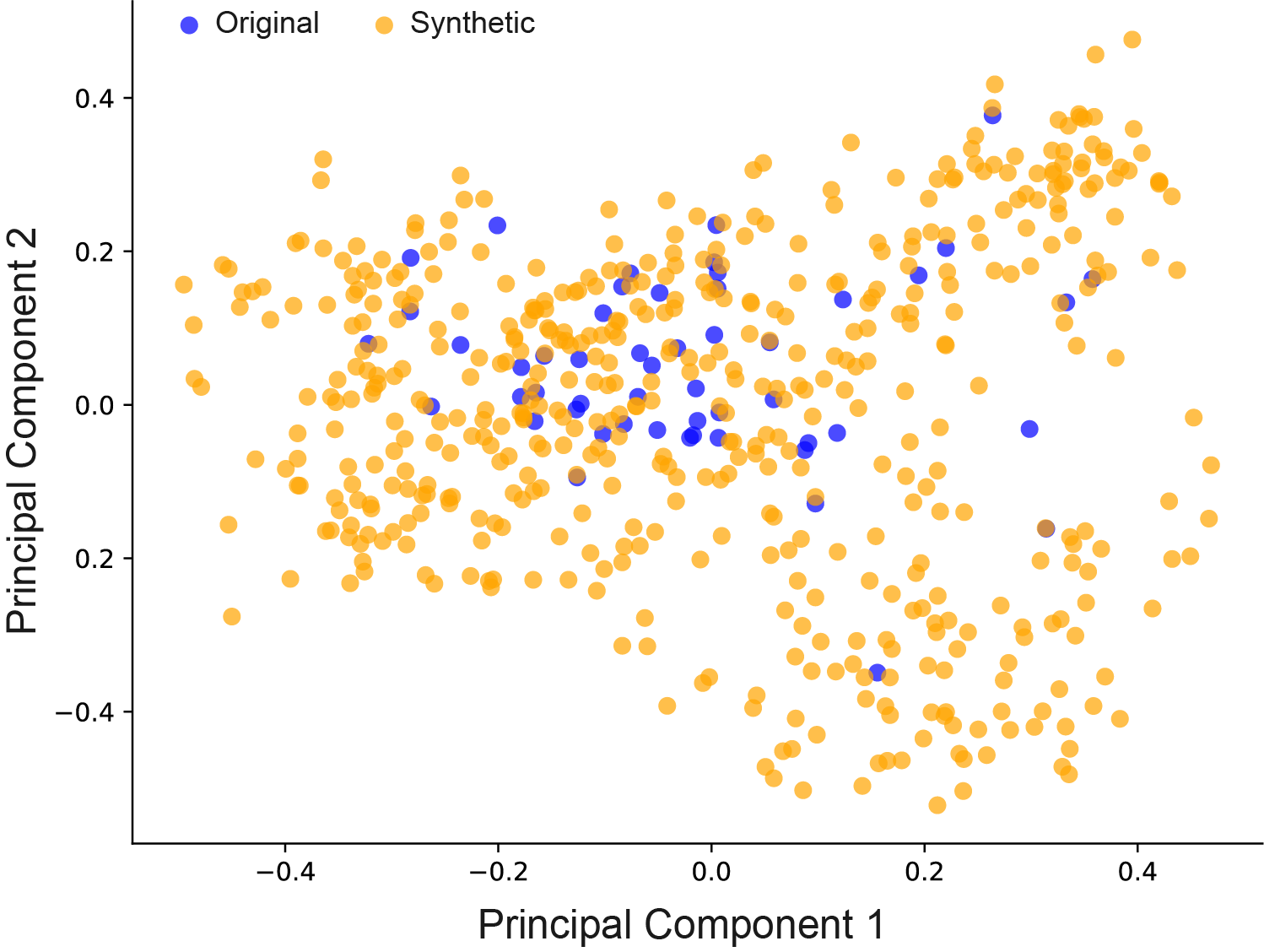}
    \caption{PCA visualization of sentence embeddings generated by Sentence Transformers (SBERT) for original (blue) and synthetic (orange) queries. The overlapping distributions indicate that the synthetic queries in the STAQ dataset capture semantic patterns similar to those in the original data.}
    \Description{A scatter plot showing the Principal Component Analysis (PCA) of sentence embeddings for original and synthetic queries. The x-axis represents Principal Component 1 and the y-axis represents Principal Component 2. Original queries are represented by blue dots and synthetic queries by orange dots. The blue and orange dots are heavily intermixed and overlap across the chart, visually indicating that the synthetic queries have a very similar semantic distribution to the original data.}
    \label{fig:fig2}
\end{figure}

We evaluated the STAQ dataset on two key dimensions: fidelity and face validity. Fidelity refers to the extent to which the synthetic data preserves the statistical properties and latent semantic patterns of the original dataset. To assess this, we compared vector representations of synthetic and real-world data using Principal Component Analysis (PCA) applied to paragraph-level embeddings generated by a Sentence Transformers (SBERT) model. Unlike standard BERT, which is optimized for token-level tasks, SBERT uses a Siamese network architecture to produce sentence-level embeddings that capture semantic similarity more effectively \cite{reimers2019sentence}. As shown in Figure \ref{fig:fig2}, the PCA visualization demonstrated a high degree of overlap between the distributions, confirming that the synthetic queries successfully mirror the underlying semantic structure of the original data and indicating strong fidelity.

To assess face validity, we conducted a qualitative review with two domain experts in aging and technology. A random sample of 50 synthetic queries was independently evaluated by each expert. Reviewers classified each query based on how likely it was to have been communicated by an older adult, using the following categories: (1) likely, (2) possibly, or (3) unlikely. Of the 50 queries, 40 were rated as `likely,' indicating strong plausibility (nine were rated `unlikely', one `possibly'). Inter-rater reliability was high (Cohen’s $\kappa = 0.83$). These results support the face validity of the STAQ dataset as a realistic representation of how older adults communicate when seeking technology support.

\section{Discussion}
Despite AI’s potential to enhance older adults’ well-being and independence, it also risks reinforcing digital ageism \cite{neven2017triple}. This bias may widen the digital divide not just between generations, but also within older populations along lines of education, income, and access \cite{friemel2016digital,gell2015patterns}. Although device ownership is rising, many older adults still struggle with tasks beyond basic use \cite{AARPTechTrends2026,czaja2019usability}. Mainstream AI systems, often designed for clear and concise inputs, may exclude users whose communication styles are different.

This work focuses on technology support, which lies at the intersection of growing need and increasing automation. While older adults require help with continuously evolving digital systems \cite{lazar2026interrogating}, support is increasingly delegated to AI-based agents. These systems often assume users can engage with advanced tools at the very moment they experience confusion—risking what might be called algorithmic ageism \cite{wu2024atlas}.

As a result, technology support often falls to family and friends, who adopt caregiving roles long before physical support is needed \cite{sharifi2023cross}. This  ``technology caregiving'' also presents challenges, especially when older adults struggle to articulate problems. Queries that are vague, overly detailed, or use unfamiliar terms can be hard to interpret. This study deepens our understanding of these barriers and introduces a age-friendly approach to make support systems more inclusive.

\subsection{Primary Findings}
Our study offers empirical insight into how older adults communicate technology problems in unstructured contexts. We documented how help-seeking queries are often incomplete, vague, or overly detailed (Table ~\ref{tab:queries}), directly conflicting with current AI tools that assume contextual information is already available \cite{gao2024easyask,wang2023enabling}. To address this, we developed an LLM–based approach that mimics a human helper: asking contextual questions, paraphrasing the query, and generating a solution. Using few-shot prompt chaining, we found that GPT-4o effectively reformulated unstructured queries, substantially increasing the retrieval of correct solutions compared to the original queries. Importantly, this LLM-mediated articulation benefited both sides of the support interaction. LLM-rephrased queries were more easily understood by younger adults, who often serve as informal technology helpers, improving their comprehension and confidence. Simultaneously, older adults evaluated the LLM-generated solutions as easy to understand and follow. These findings highlight the potential for LLMs to reduce the communicative burden of help-seeking and support diverse aging populations.

The specific magnitudes we report—the solution-accuracy and comprehension gains from query rephrasing—are a snapshot of current models. As language models continue to improve, we anticipate that future models will handle unstructured input better and may supersede our specific workflow. Our contribution, however, is not the LLM pipeline \textit{per se}, but the knowledge of how older adults articulate technology problems, where the breakdown happens, and how LLMs can mitigate it. Our experiments further demonstrate that mitigating these breakdowns improves technology support on both sides of the interaction—for the older adults receiving it and the technology helpers providing it.

\subsection{Limitations}

Our findings should be interpreted in light of several limitations. First, the older adult participants across our diary study and evaluation experiments comprised a relatively small, self-selected group of community-dwelling, English-speaking individuals. All possessed baseline experience with smart devices, meaning our sample lacked digital novices. Consequently, while their queries reflect common frustrations, these findings may not generalize to users with minimal technology exposure or significant cognitive impairments. Second, while our formative diary study captured naturalistic help-seeking, our subsequent evaluation experiments were conducted in controlled settings; rating static queries and solutions may not fully reflect the dynamic, conversational complexities of real-world support interactions. Finally, our approach relied on a state-of-the-art, proprietary large language model (GPT-4o). This performance may not be replicable with smaller or open-source models, potentially limiting scalability in resource-constrained environments. Furthermore, reliance on a proprietary API means that future changes to model behavior could affect output consistency and reliability.

\subsection{Implications for Aging Research and Practice}
This work advances aging research by advancing understanding of how older adults articulate technology problems—an often-overlooked aspect of aging with technology. While prior studies have largely focused on whether older adults adopt certain technologies or why they may resist them \cite{czaja2006factors}, our study shifts attention to what happens after adoption—when older adults encounter difficulties and seek help. The communication patterns we identified offer a vocabulary for describing common pitfalls in help-seeking, which can inform both research and practice.

Digital literacy programs, for instance, could incorporate modules that teach older adults how to formulate more effective help requests—such as how to emphasize the core issue and provide relevant details. However, it may be unrealistic to expect all older adults to adopt concise, technically precise language, particularly given age-related cognitive changes and rapidly changing technology. This reinforces the need for support systems that adapt to users, rather than forcing users to adapt to systems.

Our LLM-based approach offers a model for such age-friendly design. By beginning with real-world diary entries and designing around older adults’ actual communication styles—rather than defaulting to younger user norms—we treat older adults not as deficient, but as capable users with distinct interaction patterns. Our diary study revealed a strong reliance on text, with only a small fraction of queries including screenshots---and none utilizing screen or video recording options. This limited use of multimodal context may stem from the cognitive overhead required to capture and attach visual information, unfamiliarity or difficulty with those actions, or underlying privacy concerns. Future iterations of the pipeline could address this barrier by dynamically generating instructions during the follow-up phase, proactively guiding older adults through capturing and sharing screenshots when visual context is necessary.

AI-powered technology support could also extend the reach of existing technology  education efforts. For example, community centers or libraries could deploy AI chatbots to provide just-in-time assistance for older patrons—especially those without digitally skilled family members or immediate access to help. These systems could foster autonomy, allowing older adults to solve problems on their own schedules rather than waiting for in-person assistance. While our evaluation provides a cross-sectional view of help-seeking, longitudinal deployments are necessary to understand how such a system impacts problem articulation over time. Observing repeated interactions with the system could reveal whether the tool exerts a pedagogical effect, gradually training older adults to independently formulate more structured and complete technology support queries.

Finally, our study emphasizes the diversity within the older adult population—across cognitive abilities, digital proficiency, and communication styles. Some individuals prefer narrative, detailed explanations; others are brief and to the point. Our LLM pipeline was able to handle both over- and under-specified queries, suggesting its potential to support a range of help-seeking behaviors. For aging services practitioners, this flexibility enables more tailored support. AI tools could triage requests by prompting more detail from minimal input or distilling key information from verbose descriptions. In both cases, the experience becomes more personalized.

\subsection{Implications for AI Research and Practice}
Our findings have broad implications for the future of AI research and practice, particularly in the development of human-centered systems that serve diverse populations. In addressing the technology support needs of older adults, our study highlights a persistent disconnect between users’ real-world help-seeking behaviors and the assumptions built into many current AI systems. Most AI models are trained on data from younger, digitally fluent populations, resulting in interaction norms that often misalign with older adults’ communication styles. These biases can lead to systems that misinterpret older users’ queries or produce responses that require digital proficiency many do not possess.

Our study challenges these assumptions by showing that large language models (LLMs) can be adapted to accommodate older adults’ natural language through targeted prompt engineering. By centering the real-world communication patterns of older adults, we demonstrate that inclusive design can lead to more accessible and effective support experiences—not only for older users, but for everyone. Clarity, context sensitivity, and adaptability are not niche accommodations; they are universal design principles for equitable AI.

While this work focuses on characterizing and supporting older adults’ problem articulation, future research should explore model-centric applications for STAQ. To our knowledge, it is the first structured dataset focused specifically on this form of help-seeking behavior. STAQ can serve as an auditing benchmark to evaluate whether commercial LLMs exhibit systematic performance drops on older adult-style queries compared to standard technical support corpora. Comparing performance on STAQ against standard, crowd-sourced data (which typically skew younger) can reveal algorithmic disparities and directly inform the refinement of inclusive models. However, deploying LLM-assisted pipelines in real-world technology caregiving contexts necessitates a critical evaluation of privacy and trust. Processing potentially sensitive, personal technology queries through cloud-based APIs introduces privacy concerns for both older adults and their helpers, underscoring the need to explore secure, localized, or on-device model processing in future system designs.

This work also surfaces key methodological insights for the AI research community. First, we demonstrate that synthetic datasets can successfully capture users’ lived experiences when grounded in empirical, user-centered data, offering a scalable path toward inclusive design. However, as prompt-based data generation becomes increasingly popular, it inherently risks producing stereotypical or hallucinated outputs if not anchored to real user needs. We mitigated this by strictly integrating domain-informed and user-centered strategies into our generation pipeline. As synthetic generation expands, researchers must continue to balance computational scalability with rigorous, human-centered evaluation frameworks to ensure AI tools reflect true real-world diversity rather than algorithmic assumptions.
\section{Conclusion}

When seeking support to resolve technology problems, older adults typically either search for answers online or turn to a trusted person in their social network. In either case, they must articulate the problem effectively enough to reach a resolution. In this paper, we identified the communication challenges older adults face when describing technology problems and examined how these challenges affect both the quality of support they receive and their ability to find solutions independently (e.g., via a Google search).

To address these challenges, we developed an LLM pipeline that reformulates unstructured queries into clearer, more actionable requests and generates targeted solutions---reducing the cognitive burden on technology helpers and improving the likelihood of successful independent resolution through online resources. To support further research in this space, we introduced STAQ, a synthetic dataset of older adults' technology problems designed to enable more realistic evaluation of AI-assisted support systems.

Ultimately, this work emphasizes that older adults are not deficient users, but communicators with distinct interaction styles that current AI systems often overlook. Embedding these authentic styles into AI training, evaluation, and deployment processes is essential for building equitable, age-inclusive technologies.

\section*{Acknowledgments}
This work was supported in part by the National Institute on Aging of the National Institutes of Health under Award No. P30AG083255 to DC. We thank our study participants for their time and Tasneem Mubashshira for her assistance with data analysis.

\section*{Ethical Considerations}
All study procedures were approved by the university’s Institutional Review Board (IRB). A secure version of ChatGPT \cite{achiam2023gpt} was used to assist with language editing and article formatting. No AI-generated content was included without human review, revision, and approval. The authors retain full responsibility for the accuracy and integrity of the manuscript.

\bibliographystyle{ACM-Reference-Format}
\bibliography{bib}
\clearpage
\appendix
\section{LLM-Generated Paraphrasing and Solutions
}
\label{app:a}
In this section, we provide detailed supplementary material for the prompts and examples used to generate contextual questions, paraphrased queries, and solutions. 

\subsection{Prompts used with GPT-4o
}
\begin{figure*}[h]
    \centering
    \includegraphics[width=\linewidth]{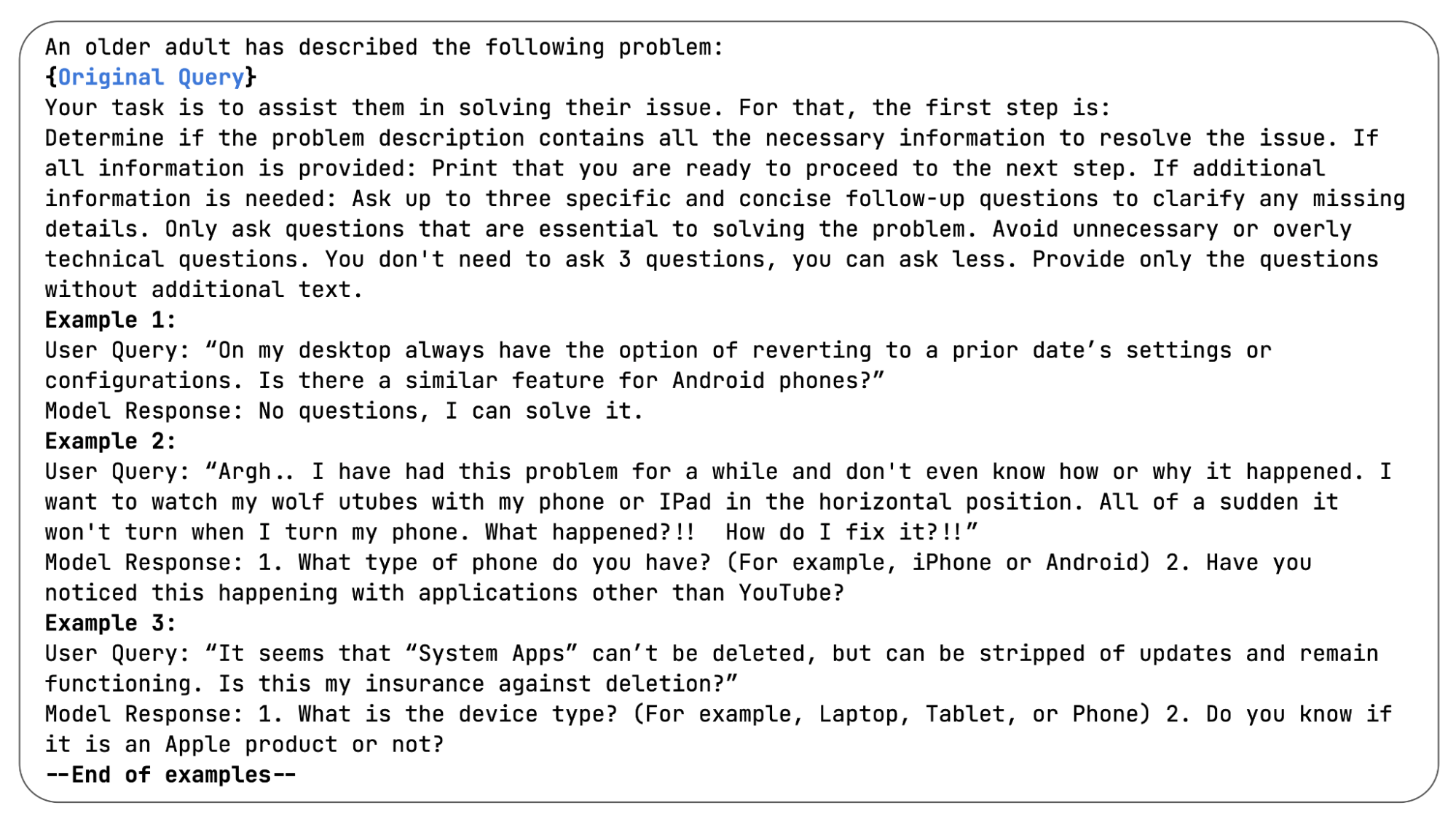}
    \caption{The prompt used for generating contextual questions.}
    \Description{A screenshot of the system prompt used to generate contextual questions. The prompt instructs the LLM to determine if a user's problem description contains all necessary information. If information is missing, the LLM is instructed to ask up to three specific and concise follow-up questions to clarify details, avoiding overly technical language. The prompt includes three few-shot examples demonstrating user queries and the expected model responses.}
    \label{fig:3}
\end{figure*}

\begin{figure*}
    \centering
    \includegraphics[width=0.9\linewidth]{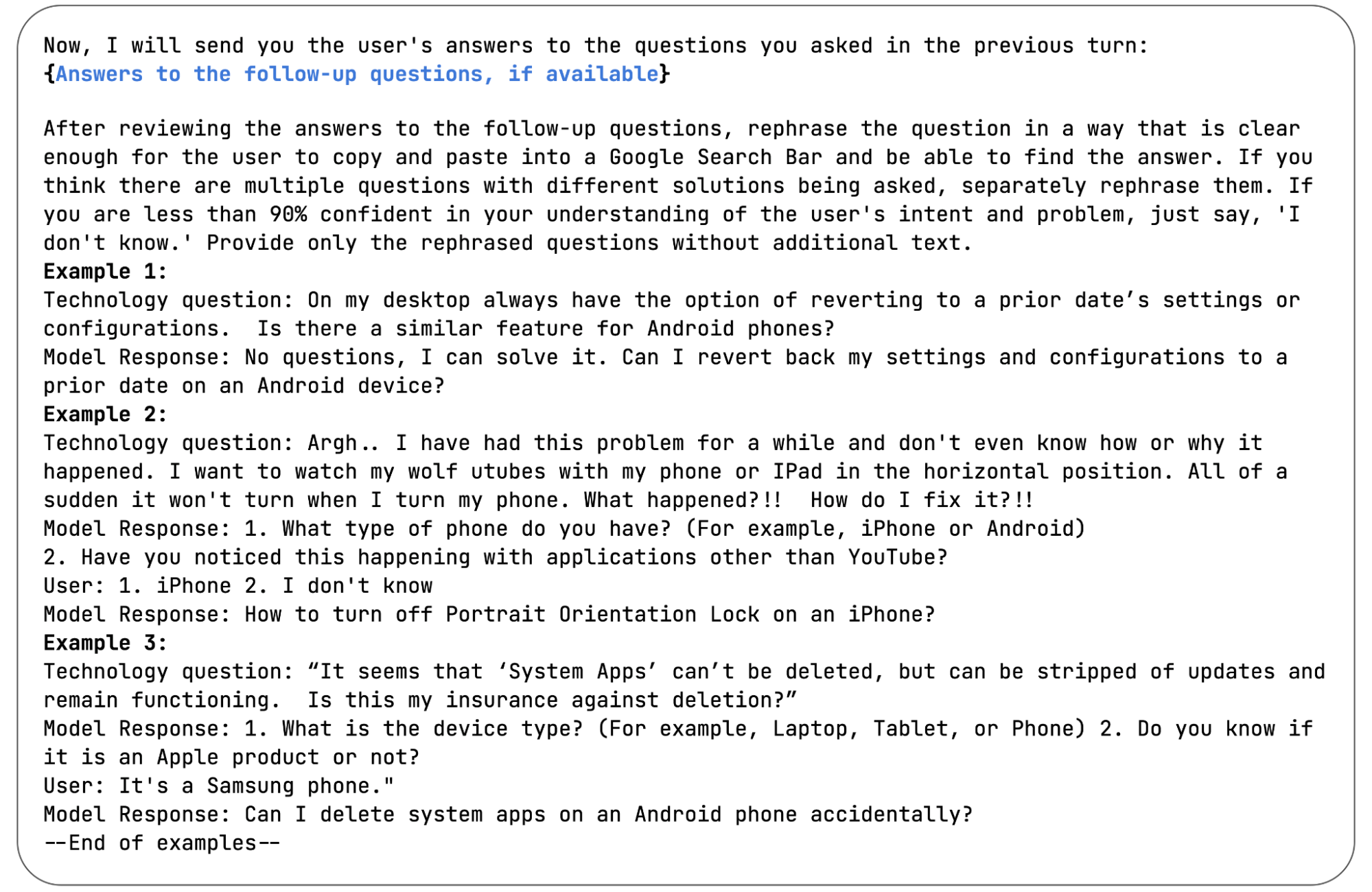}
    \caption{The prompt used for paraphrasing a technology query.}
    \Description{A screenshot of a system prompt used to paraphrase queries. It instructs the model to take a user's answers to follow-up questions and rephrase the original technology question so it is clear enough to be copied and pasted into a Google Search bar. It includes three few-shot examples showing the original technology question, the user's answers to follow-up questions, and the resulting clear, paraphrased query.}
    \label{fig:4}
\end{figure*}

\begin{figure*}
    \centering
    \includegraphics[width=\linewidth]{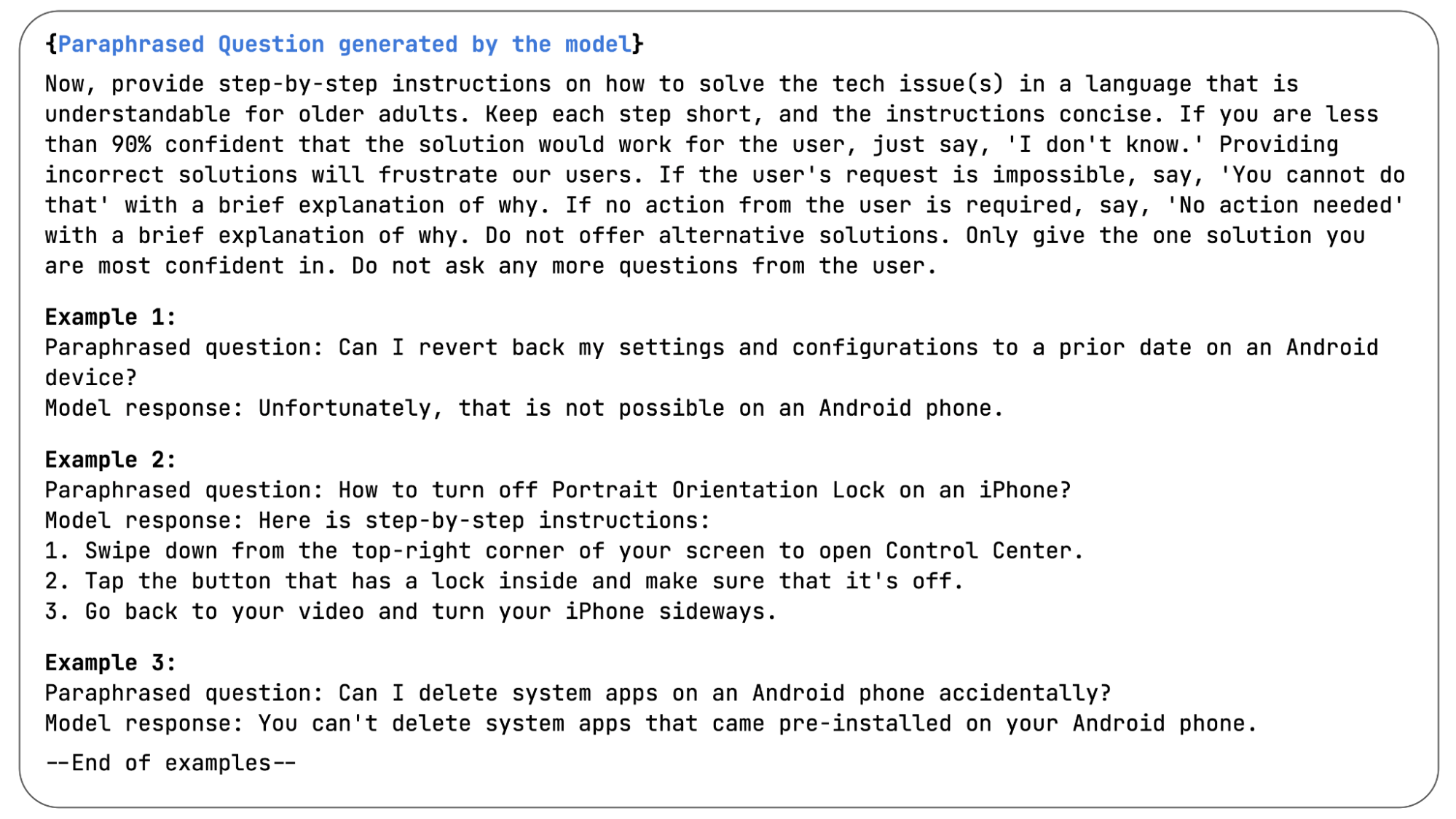}
    \caption{The prompt used for generating a technology solution.}
    \Description{A screenshot of a system prompt instructing the model to provide step-by-step instructions for an older adult based on a paraphrased technology question. It instructs the LLM to keep steps short, state 'I don't know' if confidence is below 90 percent, and not offer alternative solutions. It includes three few-shot examples of paraphrased questions and the expected step-by-step model responses.}
    \label{fig:5}
\end{figure*}

Figures \ref{fig:3},\ref{fig:4}, and \ref{fig:5} show the prompts corresponding to contextual question generation, query paraphrasing, and solution generation, respectively.

\subsection{Examples of LLM-Generated Paraphrasing }
\begin{figure*}[h]
    \centering
    \includegraphics[width=\linewidth]{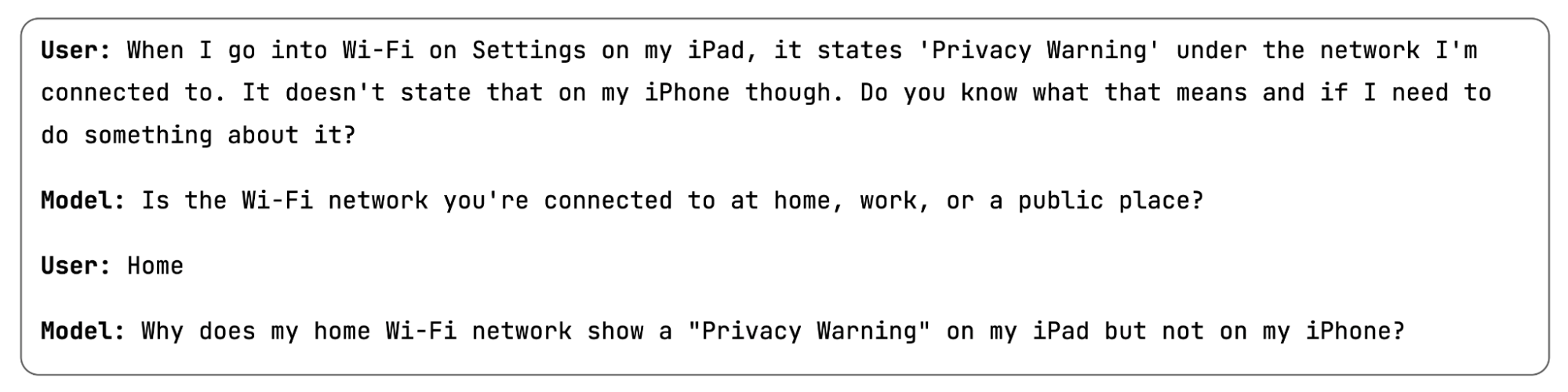}
    \caption{Example of a correct paraphrasing.}
    \Description{A transcript showing a successful query paraphrasing. The user asks about a 'Privacy Warning' on their iPad Wi-Fi. The model asks if the network is at home, work, or public. The user replies 'Home'. The model successfully combines this context to output the paraphrased query: "Why does my home Wi-Fi network show a 'Privacy Warning' on my iPad but not on my iPhone?"}
    \label{fig:6}
\end{figure*}

\begin{figure*}[h]
    \centering
    \includegraphics[width=\linewidth]{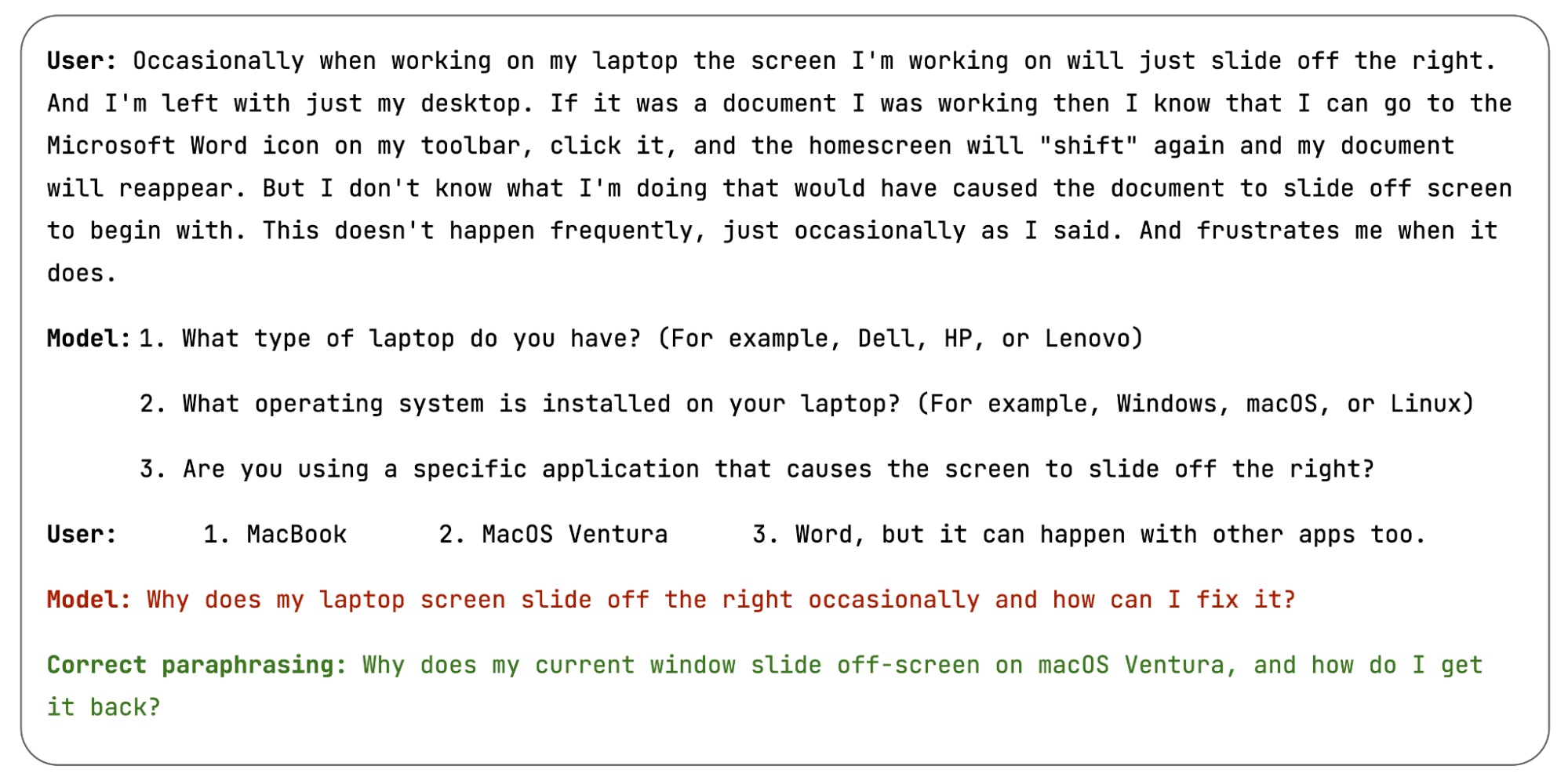}
    \caption{Example of an incorrect paraphrasing.}
    \Description{A transcript showing an incorrect paraphrasing. The user describes a laptop screen occasionally sliding off to the right. The model asks for the laptop brand, operating system, and specific app. The user replies: MacBook, MacOS Ventura, and Word. The model outputs a generic paraphrase: "Why does my laptop screen slide off the right occasionally and how can I fix it?", failing to incorporate the specific operating system. The correct expert paraphrase is shown below it, which successfully includes "macOS Ventura".}
    \label{fig:7}
\end{figure*}

Figures \ref{fig:6} and \ref{fig:7} illustrate cases where the model produced correct and incorrect paraphrases, respectively.

\subsection{Examples of LLM-Generated Solutions
}
\begin{figure*}[h]
    \centering
    \includegraphics[width=\linewidth]{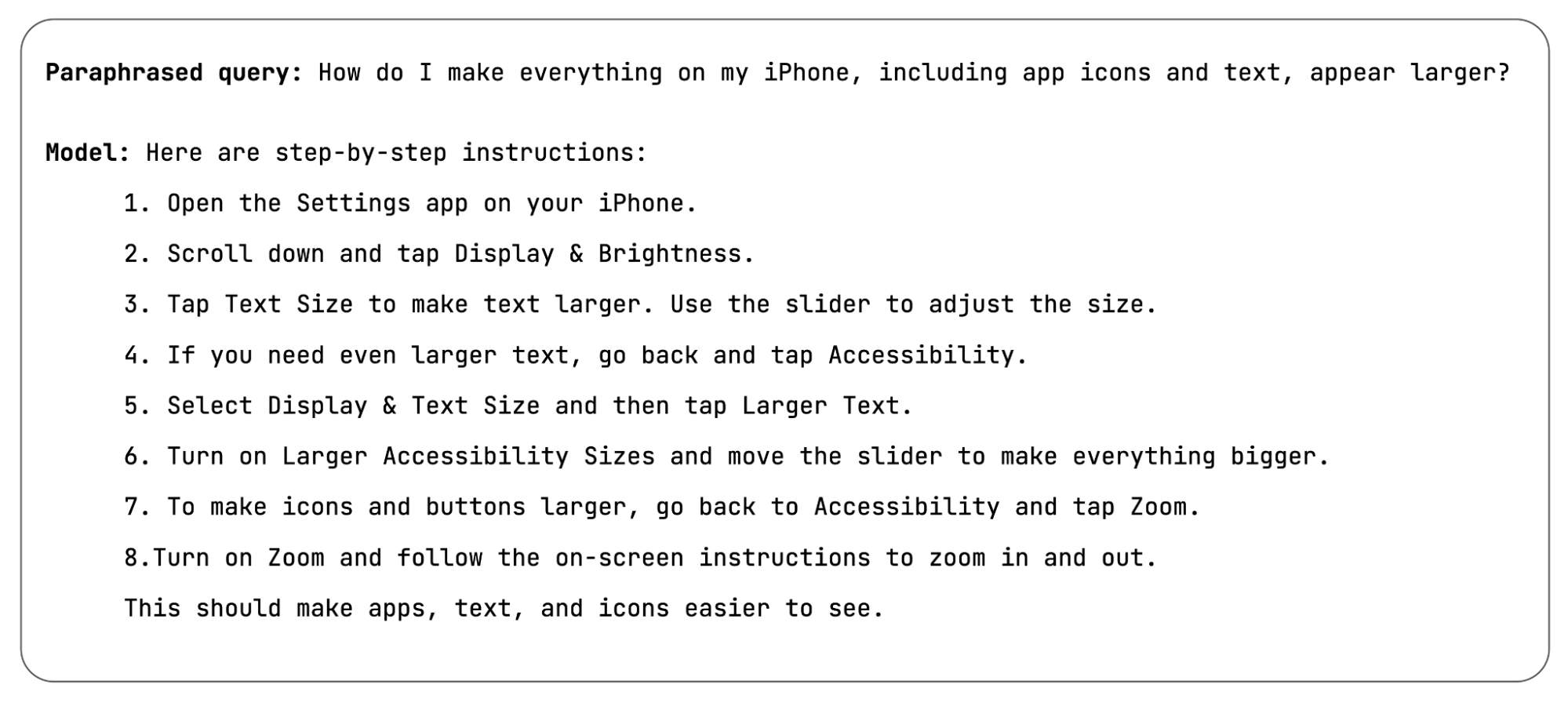}
    \caption{Example of a correct technology solution.}
    \Description{A transcript showing a correctly generated technology solution. For the paraphrased query "How do I make everything on my iPhone, including app icons and text, appear larger?", the model outputs an accurate eight-step guide detailing how to navigate the 'Display and Brightness' and 'Accessibility' menus in the iPhone Settings app.}
    \label{fig:8}
\end{figure*}

\begin{figure*}[h]
    \centering
    \includegraphics[width=\linewidth]{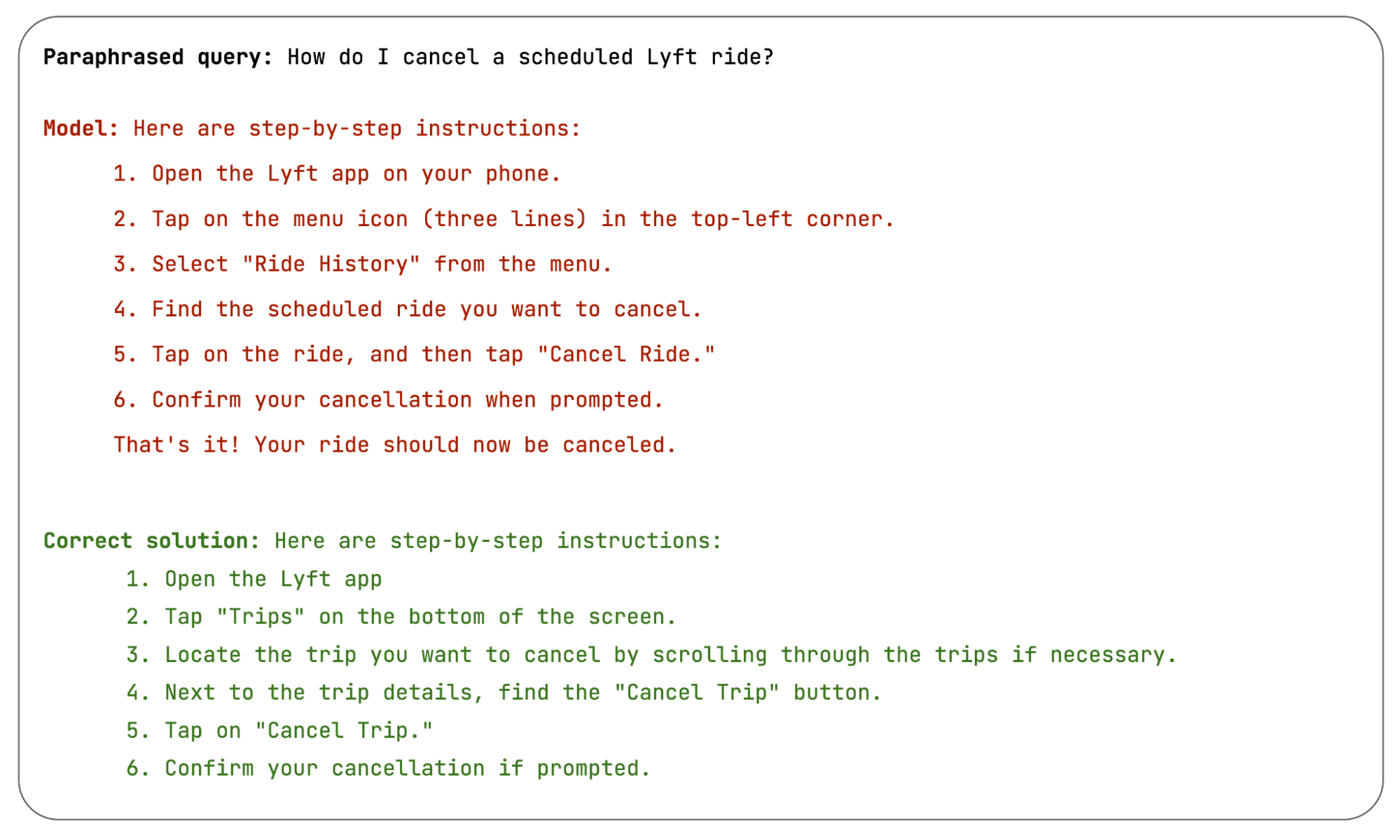}
    \caption{Example of an incorrect technology solution.}
    \Description{A transcript showing an incorrectly generated technology solution for canceling a scheduled Lyft ride. The model's instructions incorrectly suggest tapping a menu icon and selecting "Ride History". Below it, the correct expert solution is provided, which accurately instructs the user to tap "Trips" at the bottom of the screen to locate and cancel the ride.}
    \label{fig:9}
\end{figure*}

Figures \ref{fig:8} and \ref{fig:9} show correct and incorrect solution outputs.
\subsection{Quantitative Evaluation of Model Outputs}
To complement our qualitative analyses, we conducted a quantitative evaluation of model-generated paraphrases and solutions using established natural language processing (NLP) metrics. Our goal was to assess how closely the LLM-generated content aligned with expert-written references, especially in the context of older adults’ help-seeking behaviors. The following measures were used.

BLEU \cite{papineni2002bleu} is a precision-based metric originally developed for evaluating machine translation. It measures how many words or short sequences of words (n-grams) in the model output match those in a reference text. Higher scores indicate greater similarity to the reference. In this study, BLEU was used to assess how well LLM-generated paraphrases and solutions matched expert-written versions.

ROUGE-L \cite{lin2004rouge} measures the longest common subsequence between the model output and a reference text, capturing both precision and recall. It is often used to evaluate the quality of summaries. Here, it was used to assess whether model-generated solutions preserved the key steps and sequence of expert solutions.

BERTScore \cite{zhang2019bertscore} compares the contextual embeddings of words in the model output and reference text using a pretrained language model (BERT). It goes beyond exact word matches to assess semantic similarity. Higher scores suggest closer alignment in meaning, even if different wording is used.

Cosine similarity measures the angle between two vectors—in this case, the semantic representations of the model output and reference text. It ranges from $-1$ to $1$, where 1 indicates identical semantic orientation. This metric was used to quantify how similar the model-generated solutions were to the expert-written solutions at a conceptual level.

ParaScorer \cite{shen2022evaluation} is a paraphrase evaluation framework that combines reference-based and reference-free metrics. Reference-based metrics compare model output to an expert-written paraphrase, while reference-free metrics assess semantic similarity and fluency without relying on a reference. This tool helps evaluate how well LLM systems reformulate user queries for improved clarity and problem-solving effectiveness.

Paraphrasing quality was evaluated using ParaScorer \cite{shen2022evaluation}, which provides both reference-based and reference-free metrics. Reference-based metrics assess the degree of similarity between model outputs and expert-generated paraphrases, while reference-free metrics evaluate semantic similarity and fluency without a predefined reference. GPT-4o outperformed OS-ATLAS on reference-based evaluation (0.95 vs. 0.85), indicating stronger alignment with expert rewordings. In contrast, OS-ATLAS performed better on reference-free evaluation (0.70 vs. 0.60). However, paraphrasing older adults’ queries often requires adding contextual information not present in the original message, which increases lexical divergence and may penalize models in reference-free evaluation. For our use case—clarifying real-world, underspecified queries—reference-based alignment was a more relevant indicator of effectiveness. Accordingly, GPT-4o’s performance suggests greater suitability for supporting older adults through LLM-powered rewording of technology queries.

To further assess paraphrasing quality, we compared model outputs to expert-generated paraphrases using widely accepted natural language processing metrics. These included contextual similarity \cite{zhang2019bertscore}, machine translation quality \cite{papineni2002bleu}, summarization accuracy \cite{lin2004rouge}, and cosine-based semantic similarity. As shown in Table \ref{tab:paraquant}, GPT-4o consistently outperformed OS-ATLAS across all metrics. We applied these same metrics to evaluate model-generated solutions against expert-written responses (Table \ref{tab:sol-eval}). GPT-4o again demonstrated stronger alignment, achieving a BLEU score of 0.62 and a ROUGE-L score of 0.72, compared to OS-ATLAS scores of 0.04 and 0.30, respectively. Because BLEU and ROUGE measure n-gram overlap and sequence fidelity, they are particularly well suited for evaluating multi-step technology support instructions—where both accuracy and clarity of sequence are essential for usability. This is especially important for older users, who often benefit from step-by-step, easy-to-follow guidance when resolving technology issues.

\begin{table}[h]
\centering
\caption{Quantitative evaluation of query paraphrasing.}
        \label{tab:paraquant}
        \centering
        \begin{tabular}{ccccc}
            \toprule
            \textbf{Model} & \textbf{BERTScore} & \textbf{BLEU} & \textbf{ROUGE-L} & \textbf{Cosine} \\
            \midrule
            GPT-4o & .98 & .75 & .87 & .88 \\
            OS-ATLAS & .92 & .09 & .40 & .36 \\
            \bottomrule
            \\
        \end{tabular}
    \end{table}

\begin{table}[h]
    \centering
       \caption{Quantitative evaluation of solutions generated.}
        \label{tab:sol-eval}
        \centering
        \begin{tabular}{ccccc}
            \toprule
            \textbf{Model} & \textbf{BERTScore} & \textbf{BLEU} & \textbf{ROUGE-L} & \textbf{Cosine} \\
            \midrule
            GPT-4o & .94 & .62 & .72 & .79 \\
            OS-ATLAS & .87 & .04 & .30 & .38 \\
            \bottomrule
        \end{tabular}
\end{table}



\section{Study Instruments
}
\label{app:b}

How much do you agree or disagree with the following statements? (5-point scale: 1 = Strongly Disagree, 5 = Strongly Agree)
\begin{itemize}
    \item I know how to protect others and my own personal data (e.g., photos, addresses, family information) online.
    \item I can verify online information using multiple sources.
    \item I understand what hardware and software technologies are.
    \item I can install software on my computer or other electronic devices.
    \item I understand the terms ``Internet" and ``World Wide Web (WWW)."
    \item I can effectively use e-government applications, such as online tax filing, driver's license applications, and voter registration.
    \item I can use cloud computing technologies (Google Drive, iCloud, Dropbox, etc.) effectively in daily life.
    \item I can use the calendar on mobile devices not only just to look at dates but also to check dates, set reminders, and create events.
    \item I can upload and broadcast videos online.
    \item I can use digital technologies effectively in daily practice such as reservation, shopping, address finding, etc.
   \item I can add a web page that I use to bookmarks or favorites. 
    \item I know how to restrict apps' access to my personal information (location, contacts, camera, etc.)
    \item I can recognize and block unwanted / spam emails and phishing messages. 
    \item I can change the privacy/security settings on my social media posts and profile.
    \item I know how to create a strong and secure password.
    \item I can effectively use at least one software program related to my field (e.g., Photoshop, SPSS, Premiere, Microsoft Word).
\end{itemize}
The final digital literacy score was computed using the mean of all ratings \cite{bayrakci2021digital} (Bayrakci et al., 2021).

\section{STAQ}
\label{app:c}
\subsection{Prompt used with GPT-4o
}
\begin{figure*}[h]
    \centering
    \includegraphics[width=\linewidth]{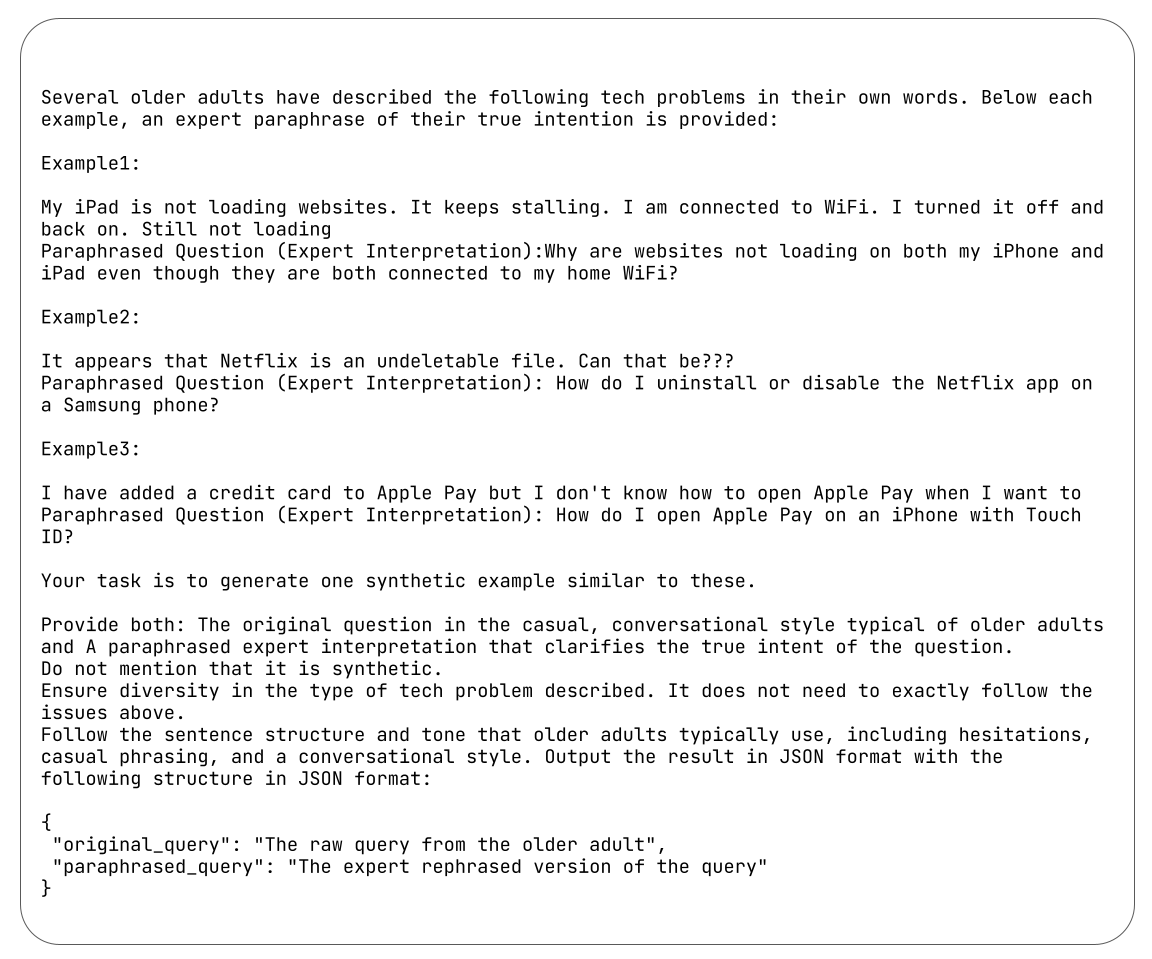}
    \caption{Prompt used to generate synthetic data.}
    \Description{A screenshot of the system prompt used to generate synthetic queries for the STAQ dataset. The prompt provides three examples of older adults' original, conversational tech problems paired with expert paraphrased versions. It instructs the LLM to generate one new synthetic example mimicking the casual, conversational style, hesitations, and sentence structure typical of older adults, and to output the result in JSON format.}
    \label{fig:10}
\end{figure*}

Figure \ref{fig:10} shows the prompt used to generate synthetic data with few-shot examples.

\end{document}